\documentclass[10pt,3p]{elsarticle}
\usepackage{stix2}
\usepackage{amsmath}
\usepackage{amsthm}
\usepackage{graphicx}
\usepackage{epstopdf} 
\usepackage{bm}
\usepackage{todonotes}
\usepackage{booktabs}
\usepackage{caption}
\usepackage{subcaption}
\usepackage{hyperref}
\usepackage{xurl}
\usepackage{placeins}
\usepackage[mathlines]{lineno}
\usepackage{comment}
\usepackage{cancel}
\usepackage{cleveref}
\graphicspath{{figures/}}

\newcommand{\revtwo}[1]{\textcolor{black}{#1}}

\makeatletter
\def\ps@pprintTitle{%
 \let\@oddhead\@empty
 \let\@evenhead\@empty
 \def\@oddfoot{}%
 \let\@evenfoot\@oddfoot}
\makeatother

\newcommand*\linenomathpatchAMS[1]{%
  \expandafter\pretocmd\csname #1\endcsname {\linenomathAMS}{}{}%
  \expandafter\pretocmd\csname #1*\endcsname{\linenomathAMS}{}{}%
  \expandafter\apptocmd\csname end#1\endcsname {\endlinenomath}{}{}%
  \expandafter\apptocmd\csname end#1*\endcsname{\endlinenomath}{}{}%
  }
\newcommand{\rd}{\mathrm{d}}
\newcommand{\vt}[1]{\bm{#1}}
\newcommand{\Rey}{\mathrm{Re}}

\newcommand{\Pran}{\mathrm{Pr}}
\newcommand{\Ray}{\mathrm{Ra}}

\newcommand{\Nu}{\mathrm{Nu}}

\expandafter\ifx\linenomath\linenomathWithnumbers
  \let\linenomathAMS\linenomathWithnumbers
  \patchcmd\linenomathAMS{\advance\postdisplaypenalty\linenopenalty}{}{}{}
\else
  \let\linenomathAMS\linenomathNonumbers
\fi

\linenomathpatchAMS{gather}
\linenomathpatchAMS{multline}
\linenomathpatchAMS{align}
\linenomathpatchAMS{alignat}
\linenomathpatchAMS{flalign}

\begin{document}
% \linenumbers

\begin{frontmatter}

\title{A pressure-free long-time stable reduced-order model \\ for two-dimensional Rayleigh-Bénard convection}

\author[1]{K. Chand}
\author[1]{H. Rosenberger}
\author[1]{B. Sanderse \corref{cor1}}
\ead{b.sanderse@cwi.nl}
\cortext[cor1]{Corresponding author}

\affiliation[1]{organization={Centrum Wiskunde \& Informatica},%Department and Organization
            addressline={Science Park 123}, 
            city={Amsterdam},
            country={The Netherlands}}

\begin{abstract}
The present work presents a stable POD-Galerkin based reduced-order model (ROM) for two-dimensional Rayleigh-B\'{e}nard convection in a square geometry for three Rayleigh numbers: $10^4$ (steady state), $3\times 10^5$ (periodic), and $6\times 10^6$ (chaotic). Stability is obtained through a particular (staggered-grid) full-order model (FOM) discretization that leads to a ROM that is pressure-free and has skew-symmetric (energy-conserving) convective terms. This yields long-time stable solutions without requiring stabilizing mechanisms, even outside the training data range. The ROM's stability is validated for the different test cases by investigating the Nusselt and Reynolds number time series and the mean and variance of the vertical temperature profile. In general, these quantities converge to the FOM when increasing the number of modes, and turn out to be a good measure of accuracy. However, for the chaotic case, convergence with increasing numbers of modes is relatively difficult and a high number of modes is required to resolve the low-energy structures that are important for the global dynamics.
\end{abstract}

\end{frontmatter}
\section*{Lead paragraph}
\textbf{Reduced order models have been introduced to accelerate the computation of turbulent fluid flows, such as  Rayleigh-Bénard convection. However, many existing reduced-order models suffer from stability problems, which limits their applicability. We propose a novel reduced-order model that preserves important symmetries of the fluid flow equations, making it suitable for integration over long periods of time. Simulations of periodic and chaotic flows show the stability of the new model, independent of its dimension.}

\section{Introduction}\label{sec:introduction}
Rayleigh-B\'{e}nard convection (RBC) is an idealized system to study natural thermal convection \citep{Benard,Rayleigh}. Natural convection is ubiquitous in nature and has applications in geophysical aspects, astronomy, and inside the planets, to name a few \citep{Ahlers_2009,Chilla_2012}. RBC is a bottom-heated and top-cooled configuration, where the flow is set in motion due to thermo-convective instabilities which arise from the thermal expansion of the working fluid. While the buoyant force destabilizes the flow (due to thermal expansion), viscous forces act as a stabilizing agent. The ratio of buoyant force to viscous force is known as the Rayleigh number $\Ray=g\beta\Delta T H^3/(\nu\alpha)$. Here, $g$ is the acceleration due to gravity, $\beta$ is the thermal expansion coefficient, $\Delta T (= T_H-T_C)$ is the temperature difference between the two isothermal plates separated by a height $H$, and $\nu$ and $\alpha$ are the viscous and thermal diffusivities, respectively. Next to the Rayleigh number, the Prandtl number $(\Pran=\nu/\alpha)$ is another control parameter (which is a fluid property). %In RBC, for a fixed aspect ratio, 

For a fixed working fluid and geometry (aspect ratio), finding the dependence of the heat flux (Nusselt number, $\Nu$) and the flow intensity (Reynolds number, $\Rey$) on the Rayleigh number are the two key issues in RBC \cite{Zhu_2017,Zhu_2018}. 
A number of theories were proposed to establish these dependencies, namely, the classical $1/3$ scaling  \citep{Malkus_1954}, the ultimate regime in turbulent convection \citep{Kraichnan_1962}, and the unified scaling theory \citep{GL_2000}. To ascertain the evidence of the existing theories and to develop new ones, a sufficiently large temporal sample of $\Nu$ and $\Rey$ is required in order to compute averages and statistics. In addition to $\Nu$ and $\Rey$, several other quantities that are of interest require simulating large time intervals. For instance, the time-average temperature field is of interest as it quantifies the mean thermal boundary layer thickness \citep{Ahlers_2009,Zhou_2010,Chand_2019} and the structure function reveals the small-scale dynamics \citep{Lohse_2010}. Furthermore, investigation of large-scale circulation and flow reversals require a large sampling interval (more than $10^4$ free-fall time units) \citep{Wang_2018}. Similarly, thermal plumes are quantified by averaging the fluctuations of vertical velocity and temperature over sufficiently large time span \citep{Emran_2012,Chand_2019}. Lastly, identification of turbulent super-structures has recently gained momentum, which are quantified by the mean temperature field \citep{Schneide_2018,Pandey_2021}. All these studies indicate the importance of long time-sampling in turbulent Rayleigh-B\'{e}nard convection. 

In numerical experiments, turbulent flows are simulated by direct numerical simulation (DNS), large-eddy simulation (LES) and Reynolds-averaged Navier-Stokes techniques (RANS), see \citep{Pope} for more details. 
For use in optimization, design, control and uncertainty quantification studies, such simulations are typically computationally too expensive, and reduced-order model (ROM) can be advantageous. There are several different ROM approaches, for instance, proper orthogonal decomposition (POD)-Galerkin methods, Krylov subspace methods, and balanced truncation \cite{antoulas2005,Benner_2015}. In the present work, we use a POD-Galerkin based ROM, where the solution is expanded in terms of POD modes and the governing equations of the full-order model (FOM) are projected onto a lower-dimensional space (the so-called Galerkin step). The lower-dimensional space is constructed from the instantaneous snapshots of the FOM. 

It has been observed that projection-based methods work appropriately for diffusion-dominated  systems, owing to the sharp decay of the singular values of the snapshot matrix \citep{Benner_2015}. In such systems, neglecting the higher modes (associated to smallest length scales, which contain little energy), makes the POD-based ROMs efficient. On the contrary, for convection-dominated problems like turbulent flows, kinetic energy dissipation occurs at the smallest length scale (Kolmogorov length scale). In this case, discarding the highest modes results in the incorrect dissipation rate in ROMs, which can make them inaccurate and/or unstable. In addition, another stability issue that can potentially arise is due to the inf-sup condition (compatibility between pressure and velocity spaces), which needs to be satisfied at the ROM level \cite{Stabile_2018}.

Several different methodologies have been used to tackle these stability issues, for instance, using closure models to include the dissipation \citep{Wang_2012,Cai_2019,Ahmed_2021} or using a $H^1$ norm instead of $l^2$ norm to construct a POD basis \cite{Iollo_2000}. Recently, Rocha \emph{et al.} \cite{Rocha_2023} used incremental POD with a non-intrusive operator inference (OpInf) approach to attain accurate results, albeit again limited to a very short time period (less than one free-fall time unit). 
Structure preservation is another way to handle the stability issues \cite{CARLBERG_2018,Benjamin_2020}. In a recent study, Sanderse \cite{Benjamin_2020} proposed a structure-preserving (kinetic energy-conserving) non-linearly stable reduced-order model for incompressible flows. The main idea behind the stable formulation was to discretize and project the governing equations in such a way that important symmetries of the incompressible Navier-Stokes equations are kept on the ROM level, leading to exact kinetic energy conservation (in the inviscid limit) and hence stability. This methodology was implemented for isothermal flows. In this work, this energy-stable ROM is extended to non-isothermal flow, allowing us to achieve the main goal of this article: \textit{perform long-time sampling with a ROM for a thermal convection problem}. Although this approach does not directly solve the issue of representing the smallest scales and associated dissipation, the stable nature of the model makes it an excellent starting point for closure model development aimed at accuracy, without being concerned about stability. 

The paper is organized as follows. In Section \ref{sec:numerical_details}, the governing equations and the numerical details of the FOM is described. Section \ref{sec:ROM} encompasses the formulation of the ROM. In Section \ref{sec:results}, stability and accuracy of the ROM are shown for three cases: steady flow \ref{case1}; periodic flow \ref{case2}; and chaotic flow \ref{case3}. Finally, the paper is summarized in Section \ref{sec:conclusion}.
% Energy-tabi at the dislity: check the work of Joseph \cite{joseph1966,joseph1976}, the book of Straughan \cite{straughan2004} (section 4.3).

\section{Full-order model description}\label{sec:numerical_details}
\subsection{Mathematical modeling}
Rayleigh-B\'{e}nard convection is a buoyancy-driven flow governed by the mass, momentum, and energy equations. It can be considered as incompressible flow by invoking the Boussinesq approximation, which allows to incorporate density variation in the body force term. The governing equations can be written in their non-dimensionalized form as: 
\begin{equation}\label{Eq:continuity}
\centering
\nabla \cdot \textbf{u}=0,
\end{equation}
\begin{equation}\label{Eq:momentum}
\centering
\frac{\partial \textbf{u}}{\partial t} +  \nabla \cdot (\textbf{u} \otimes \textbf{u}) = -\nabla p + \sqrt{\frac{\Pran}{\Ray}}\nabla \cdot (\nabla \textbf{u} + (\nabla \textbf{u})^T) + \theta \hat{\textbf{e}}_y,
\end{equation}
\begin{equation}\label{Eq:energy}
\centering
\frac{\partial \theta}{\partial t} + \nabla \cdot (\textbf{u} \hspace{1pt} \theta) = \frac{1}{\sqrt{\Pran \Ray}}\nabla^2 \theta.
\end{equation}
Here $\hat{\textbf{e}}_y$ indicates that buoyancy acts only in the vertical direction. The unknowns are $\textbf{u} = (u,v)$, $p$ and $\theta = (T - T_H)/(T_H - T_C)$, being the non-dimensional velocity, pressure, and temperature, respectively. These governing equations are obtained by non-dimensionalizing with vertical spacing $H$ between the isothermal walls, free-fall velocity $\sqrt{g\beta \Delta TH}$ and temperature difference between the hot $(T_H)$ and cold $(T_C)$ plates $\Delta T=T_H-T_C$ as length, velocity and temperature scales, respectively. 

\begin{figure}[ht!]
\centering
\includegraphics[width=0.3\textwidth]{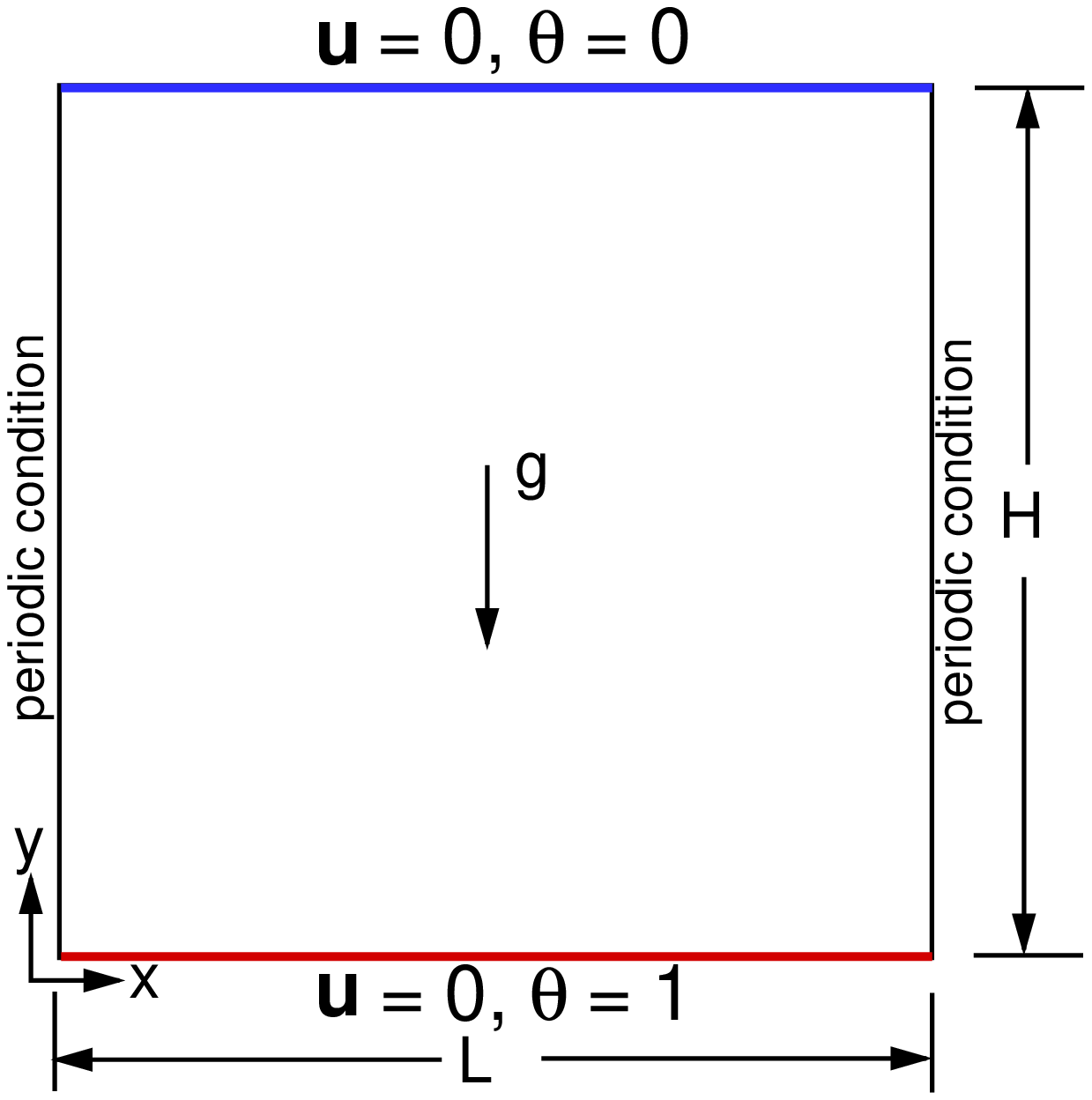}
\caption{Schematic of a $2D$ RBC case with boundary conditions.}
\label{fig:schematic}
\end{figure}

In this study, we use a $2D$ square domain filled with air as a working fluid ($\Pran=0.71$). It is worth noting that the choice of a unit aspect-ratio configuration is a commonly adopted practice when investigating the physical mechanisms within Rayleigh-Bénard convection (RBC). This applies to both two-dimensional ($2D$) and three-dimensional ($3D$) studies, see for example the following studies: \cite{Zhang_2017, Poel_2015, Cai_2019, Vishnu_2022, Li_2022}. Furthermore, the same geometric configuration with a unit aspect ratio has been recently employed by \cite{Rocha_2023}. As shown in Fig. \ref{fig:schematic}, isothermal temperature and no-slip boundary conditions are prescribed on the top and bottom walls, whereas periodic boundary conditions are used for the side walls. As initial condition, we use stagnant flow state for velocity $(u=0, v=0)$ and a linear conduction temperature profile, $\theta(x,y)=1-y/H$. To ensure the flow resolution, we carry out a grid-independent study, shown in \ref{appendix1}. Also, sufficient number of grid points constitute the boundary layer \citep{Shishkina_2010} to aptly resolve the flow near the walls, see Table \ref{tab:results_simdetails}. 

\subsection{Discrete governing equations and validation}
\begin{figure}[ht!]
\centering
\includegraphics[width=0.3\textwidth]{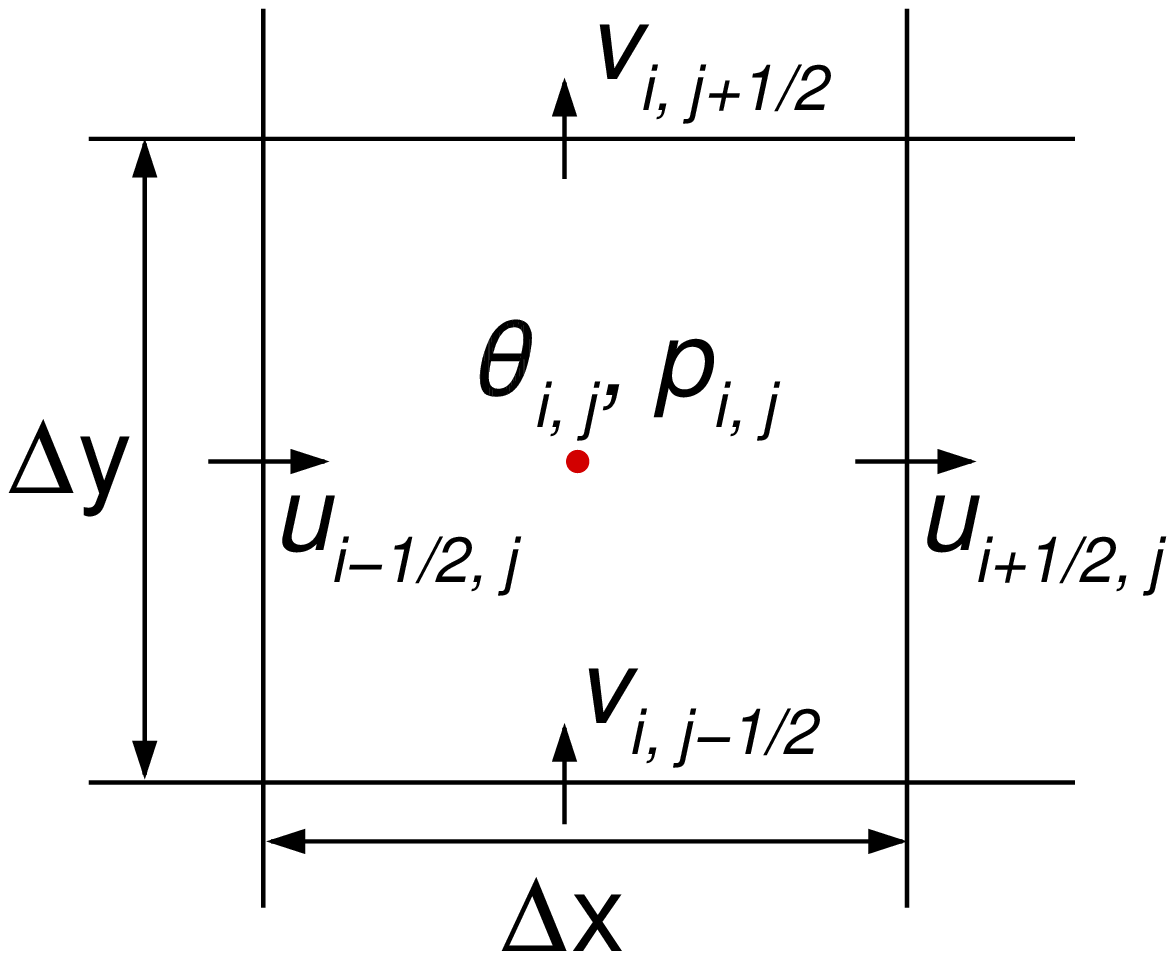}
\caption{Schematic of a $2D$ finite volume cell $p_{i,j}$ showing the location of velocity, pressure, and temperature.}
\label{fig:schm_fvm}
\end{figure}
\begin{table}[t!]
\centering
\begin{tabular}{*7c}
\toprule
Case &$\Ray$& $N_x \times N_y$ & $\Delta t$ & $N_{BL}$ & $T_{\text{sampling}}$ & NoS \\
\midrule
Steady state & $10^4$ & $48\times48$ & $10^{-2}$ &$10$ & $[0,200]$  & 20000 \\
Periodic flow&$3\times 10^5$ & $80\times80$ & $10^{-2}$ &$8$ & $[0,100]$ & 10000 \\
Chaotic flow&$6\times 10^6$ & $128\times128$ & $10^{-2}$ & $6$& $[0,50]$ & 5000 \\
   % &   & ERK  & ABCN   & ERK &ABCN\\ \cmidrule{3-6}%
\bottomrule
\end{tabular}
\caption{Simulation details. From left to right: cases, $\Ray$ is Rayleigh number; $N_x \times N_y$ are grid points in $x$ and $y$ directions, respectively, $\Delta t$ is time increment; $N_{BL}$ is number of grid points inside thermal boundary layer; $T_{\text{sampling}}$ is the temporal sampling interval; NoS is number of snapshots used to construct ROM basis.}
\label{tab:results_simdetails}
\end{table}

\begin{figure}[t]
\centering
\includegraphics[width=0.5\textwidth]{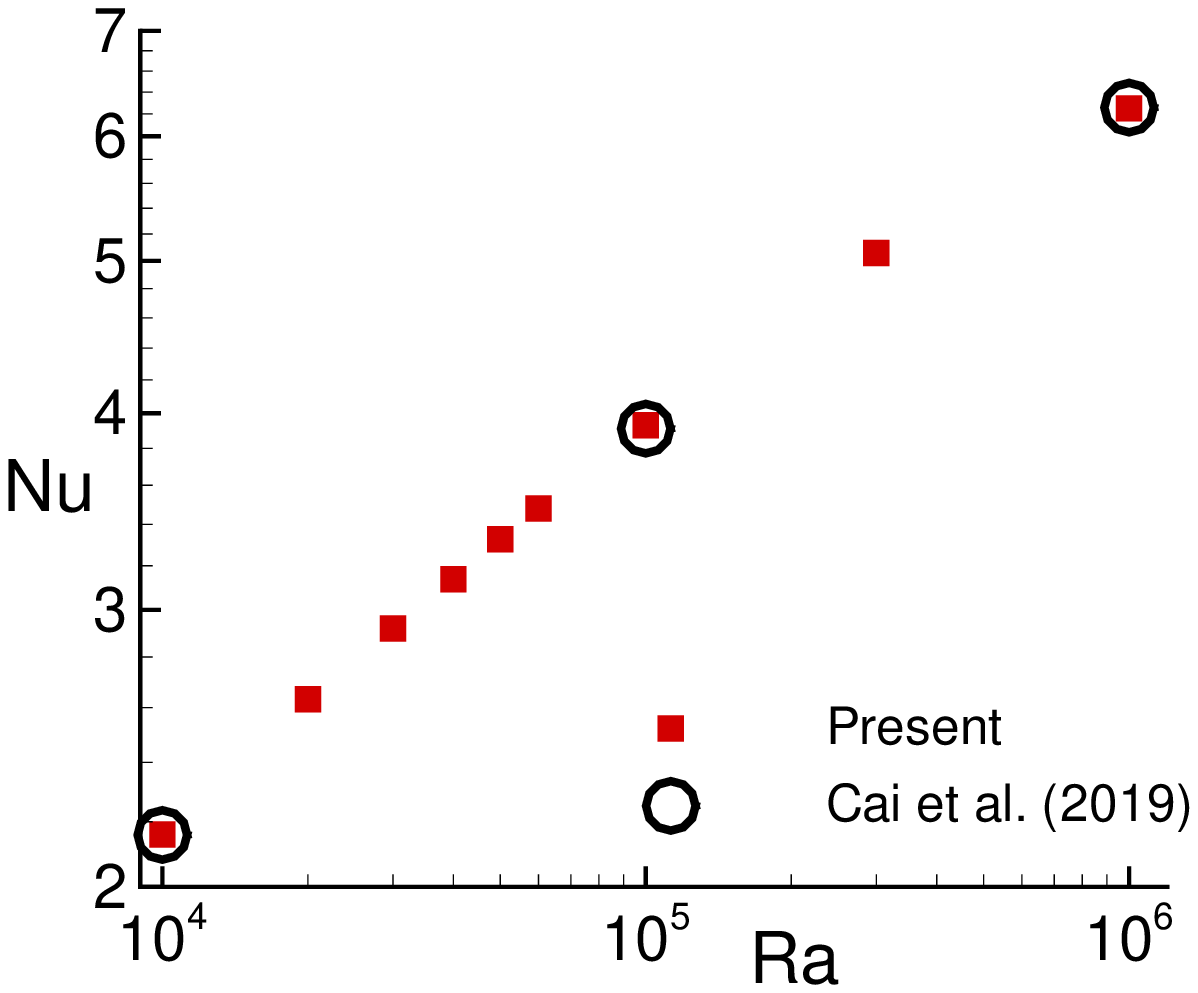}
\caption{Steady state RBC in a closed cavity: comparison of $\Nu$ obtained with our FOM to results of \cite{Cai_2019}.}
\label{fig:Nu_valid}
\end{figure} 
The governing equations \eqref{Eq:continuity}-\eqref{Eq:energy} are spatially discretized using a staggered grid finite-volume method, yielding the full-order model (FOM) that will be used to construct a reduced-order model in Section \ref{sec:ROM}. Figure \ref{fig:schm_fvm} shows a schematic of a `pressure' finite volume cell, where velocities are defined on the cell faces and pressure and temperature at the cell centre. A uniform Cartesian grid is used for all the cases. This framework has been used in our earlier work to construct stable ROMs for isothermal cases, see \cite{Benjamin_2020}. The discretized velocity, pressure, and temperature unknowns are represented by 
$V_h=\begin{pmatrix} u_h \\ v_h \end{pmatrix}$, $p_h$, and $\theta_h$, respectively. 
Note that $u_h \; (\text{horizontal velocity}) \in \mathbb{R}^{N_u}$, $v_h \; (\text{vertical velocity})\in \mathbb{R}^{N_v}$, $p_h\in \mathbb{R}^{N_p}$, and $\theta_h\in \mathbb{R}^{N_\theta}$, where $N_u$, $N_v$, $N_p$, and $N_\theta$ represent the number of finite volume cells for the unknowns mentioned in the subscript. The discretized governing equations are written as 
\begin{align}
    M V_{h}(t) &= 0, \label{eqn:mass_semidiscrete} \\
    \Omega_{V} \frac{\rd V_{h}(t)}{\rd t} &= - C_{V} (V_{h}(t)) - G p_{h}(t) +  \sqrt{\frac{\Pran}{\Ray}} D_{V} V_{h}(t) + (A  \theta_{h}(t) + y_{A}), \label{eqn:mom_semidiscrete}\\
    \Omega_{\theta} \frac{\rd \theta_{h}(t)}{\rd t} &= -  C_{\theta}(V_{h}(t),\theta_{h}(t)) + \frac{1}{\sqrt{\Pran \Ray}} (D_{\theta} \theta_{h}(t) + y_{D}).\label{eqn:energy_semidiscrete}
\end{align}
% \begin{equation}\label{Eq:continuity_dis}
%     M_h V_h = 0 
% \end{equation}
% \begin{equation}\label{Eq:momentum_dis}
%     \Omega_h \frac{d}{dt}V_h(t) = F_h^{CD}(V_h)-G_hp_h+\begin{pmatrix} 0 \\ \theta_h \end{pmatrix}
% \end{equation}
% \begin{equation}\label{Eq:energy_dis}
%     \Omega_{\theta_h} \frac{d}{dt}\theta_h(t) = F_{\theta_h}^{C_\theta D_\theta}(\theta_h)
% \end{equation}
$\Omega_V \in \mathbb{R}^{N_V \times N_V}$ and $\Omega_{\theta} \in \mathbb{R}^{N_\theta \times N_\theta}$ are diagonal matrices whose entries are the finite volume sizes of the velocity and temperature cells, respectively. $M \in \mathbb{R}^{N_p \times N_V}$ indicates the discretized divergence operator, $G = -M^T$ the discretized gradient operator, $C_V$ and $C_{\theta}$ are the (bilinear) convection operators, $D_V$ and $D_\theta$ the (linear) diffusion operators, $A$ is an averaging operator (from temperature locations to velocity locations), and $y_{A}$ and $y_{D}$ represent boundary condition contributions. For details, see \cite{sanderse2023energy}.

The discretized governing Eqs.\ \eqref{eqn:mass_semidiscrete}-\eqref{eqn:energy_semidiscrete} are solved using a fourth order explicit-Runge Kutta method, where at each stage the velocity is made divergence-free through the solution of a Poisson equation. For details, see \cite{sanderse2012accuracy}.

To establish the correct implementation of the FOM, we show a comparison of the Nusselt number $\Nu$ (computed on the bottom plate) between the present work and \citep{Cai_2019} in Fig. \ref{fig:Nu_valid}, for a range of Rayleigh numbers at which steady solutions exist. Note that the comparison is for the case of solid (no-slip) side walls, instead of the periodic ones that will be used in the time-dependent studies. For periodic boundaries, simulations become unstable (periodic or chaotic) already at lower Rayleigh numbers, as will be shown in Section \ref{sec:results}. In Fig. \ref{fig:Nu_valid}, $\Nu$ refers to the heat flux at the bottom plate, which is defined as
\begin{equation}\label{Eq:Nu}
    \Nu = -\int_{0}^L \frac{\partial \theta}{\partial y}\bigg{\lvert}_{y=0} \rd x,
\end{equation}
where $L$ represent the length of the domain. 
$\Nu$ is one of the key response parameters in RBC \citep{Ahlers_2009,Chand_2019} and is widely used to validate the numerical setup in RBC \citep{Chand_2019,Chand_2021PRF}. As shown in Fig. \ref{fig:Nu_valid}, $\Nu$ computed in the present work is in excellent agreement with the previous study \citep{Cai_2019}; see also our results in \cite{sanderse2023energy}. 

\subsection{Stability and energy conservation}\label{sec:FOM_stability}
 The choice of discrete operators plays a key role in ensuring stable-reduced order modelling \cite{Benjamin_2020}. In particular, the staggered-grid method presented in the previous section has three important properties that will make it suitable as a starting point for constructing a reduced-order model.

 The first property is that the pressure gradient term does not change the global kinetic energy $\int |\vt{u}|^2$ of the system. The pressure gradient contribution in the kinetic energy equation can be written as
 \begin{equation}
     \int_{\Omega} \nabla p \cdot \vt{u} \, \rd \Omega = \int_{\partial \Omega} p \vt{u} \cdot \vt{n} \, \rd S - \int_{\Omega} p \nabla \cdot \vt{u} \, \rd \Omega = 0,
 \end{equation}
assuming periodic or no-slip boundary conditions and $\nabla \cdot \vt{u}=0$. Our FOM inherits this property in a discrete sense:
 \begin{equation}
     V_{h}^T (G p_{h}) = -V_{h}^{T} M^T p_{h} = - (M V_{h})^{T} p_{h} = 0,
 \end{equation}
 because of the fact that the divergence and gradient operator satisfy the compatibility relation $M = -G^{T}$ on a staggered grid.

 The second property is that the convective operator is a skew-symmetric operator that does not change the global kinetic energy of the system. For the continuous equations, the convective contribution to the kinetic energy equation is
 \begin{equation}
     \int_{\Omega} \nabla \cdot (\vt{u} \otimes \vt{u})  \cdot \vt{u} \, \rd \Omega = \int_{\partial \Omega} \frac{1}{2} \|\vt{u}\|^2 \vt{u} \cdot \vt{n} \, \rd S \textcolor{green}{-} \int_{\Omega} \frac{1}{2}  \|\vt{u}\|^2 \nabla \cdot \vt{u} \, \rd \Omega = 0.
 \end{equation}
Again, the FOM inherits this property. This is most easily seen by writing the convective discretization $C_{V}(V_h)$ as $\tilde{C}_{V}(V_{h})V_{h}$ \cite{Benjamin_2020}, where $\tilde{C}_{V}$ is a skew-symmetric convective operator meaning that $\tilde{C}_{V}(V_{h}) = -\tilde{C}_{V}(V_{h})^{T}$, so that we have
\begin{equation}
    V_{h}^{T} C_{V} (V_{h}) = 0,
\end{equation}
provided that $M V_{h} = 0$ and that no-slip and/or periodic boundary conditions hold.

The third property is essentially the same as the skew-symmetry property of the convective operator in the momentum equation, but then for the convective operator in the temperature equation. The continuous equations indicate that the evolution of $\int \theta^2$ should not be changed through convection, since
 \begin{equation}
     \int_{\Omega} \nabla \cdot (\vt{u} \theta) \theta \, \rd \Omega = \int_{\partial \Omega} \frac{1}{2} \theta^2 \vt{u} \cdot \vt{n} \, \rd S {-} \int_{\Omega} \frac{1}{2}  {\theta^2} \nabla \cdot \vt{u} \, \rd \Omega = 0.
 \end{equation}
The FOM convective operator $C_\theta(V_{h},\theta_{h})$ satisfies this property, which follows again by writing it in the skew-symmetric form $\tilde{C}_{\theta}(V_{h})\theta_{h}$, with $\tilde{C}_{\theta}(V_{h}) = - \tilde{C}_{\theta}(V_{h})^{T}$ so that we have
\begin{equation}\label{eqn:skew_symmetry_theta2}
    \theta_{h}^{T} C_{\theta} (V_{h},\theta_{h}) = 0,
\end{equation}
provided that $M V_{h}=0$ and that no-slip and/or periodic boundary conditions apply for the velocity field (for details, see \cite{sanderse2023energy}).

We note that the current FOM discretization on a staggered grid is not directly suitable for non-orthogonal grids. For complex geometries, we recommend to use a discretization suitable for unstructured grids, such as the method of Trias et al.\ \cite{trias2014}. This discretization satisfies the same properties as our method (skew-symmetry of convection, div-grad compatibility), and would therefore be well-suited to serve as starting point to construct a ROM.

\section{Novel reduced-order model for natural convection flow}\label{sec:ROM}
\subsection{POD-Galerkin ROM}
In constructing a reduced-order model (ROM) for equations \eqref{eqn:mass_semidiscrete}-\eqref{eqn:energy_semidiscrete} we follow the approach for the incompressible Navier-Stokes equations presented in \cite{Benjamin_2020}, but extended to include the temperature equation. The velocity field $V_{h}(t) \in \mathbb{R}^{N_V}$ is approximated by
\begin{equation}\label{eqn:ansatz}
V_{h}(t) \approx V_{r} (t) := \Phi a(t),
\end{equation}
where $\Phi \in \mathbb{R}^{N_V \times M_{V}}$ is the POD velocity basis, $a(t) \in \mathbb{R}^{M}$ are the time-dependent velocity coefficients, and $M_{V}\ll N_{V}$. In addition to the velocity, we approximate the temperature field $\theta_{h}(t) \in \mathbb{R}^{N_p}$ by
\begin{equation}\label{eqn:ansatz_T}
\theta_{h}(t) \approx \theta_{r} (t) := \Psi b(t),
\end{equation}
where $\Psi \in \mathbb{R}^{N_{\theta} \times M_{\theta}}$ is the POD temperature basis, $b(t) \in \mathbb{R}^{M_{\theta}}$ are the time-dependent temperature coefficients, and $M_{\theta}\ll N_{\theta}$. In the test cases in Section \ref{sec:results}, we will take $M_{V} = M_{\theta} = M$\footnote{from the context the distinction between the number of modes $M$ and the discrete divergence operator $M$ should be clear}.

Equations \eqref{eqn:ansatz}-\eqref{eqn:ansatz_T} are substituted into the FOM equations \eqref{eqn:mass_semidiscrete}-\eqref{eqn:energy_semidiscrete}, and then the equations are projected by left-multiplying with $\Phi^{T}$ (momentum) and $\Psi^T$ (temperature). $\Phi$ and $\Psi$ are obtained by performing a singular-value decomposition (SVD) of snapshot matrices of velocity and temperature, such that the orthogonality conditions
\begin{equation}
\Phi^{T} \Omega_{V} \Phi = I, \qquad \Psi^T \Omega_{\theta} \Psi = I,
\end{equation}
are satisfied. The procedure to construct such bases is described in \cite{Benjamin_2020}.

This yields the ROM system of equations:
\begin{align}
\frac{\rd a(t)}{\rd t} &= - \hat{C}_{V} (a(t) \otimes a(t)) +  \sqrt{\frac{\Pran}{\Ray}} \hat{D}_{V} a(t) + (\hat{A} b(t) + \hat{y}_{A}), \label{eqn:rom_mom_semidiscrete}\\
\frac{\rd b(t)}{\rd t} &= -  \hat{C}_{\theta} (a(t) \otimes b(t)) + \frac{1}{\sqrt{\Pr \Ray}} (\hat{D}_{\theta} b(t) + \hat{y}_{D}).\label{eqn:rom_energy_semidiscrete}
\end{align}
The diffusion and buoyancy terms are linear and the associated ROM terms are easily pre-computed from $\hat{D}_{V} = \Phi^T D_{V} \Phi$, $\hat{D}_{\theta} = \Psi^T D_{\theta} \Psi$ and $\hat{A} = \Phi^{T} A \Psi$. Similarly, the boundary conditions are naturally included by the projection of the boundary vectors: $\hat{y}_{A} = \Phi^{T} y_{A}$, $\hat{y}_{D} = \Psi^{T} y_{D}$.
 The convective operators $\hat{C}_{V}$ and $\hat{C}_{\theta}$ are third-order tensors and are precomputed exactly (as alternative, one can employ energy-conserving hyper-reduction techniques such as \cite{Klein_2023}). % and are specified in \ref{sec:appendix_precompute}.
 It is important to stress that the divergence-free constraint, equation \eqref{eqn:mass_semidiscrete}, has disappeared in the ROM formulation. This is because the snapshots of the velocity are divergence-free and consequently the basis $\Phi$ is divergence-free. In addition, because of the compatibility relation $G = -M^T$, we have $\Phi^{T} G = -(M \Phi)^{T} = 0$, so the pressure gradient term in the momentum equation disappears. As a result, the ROM is pressure-free.

We note that the proposed ROM is determined by the specific form of the FOM and requires access to the different operators (convection, diffusion, pressure gradient) and associated boundary conditions. In our FOM implementation, the discretization operators are coded in terms of sparse matrix-vector operations, and building the ROM amounts to pre- and post-multiplying these matrices with the ROM basis. This constitutes a straightforward method to construct the ROM, but is also rather intrusive, and requires FOM code access. Alternatively, one can resort to non-intrusive methods like operator inference \cite{peherstorfer2016}, where the reduced operators are learned in a least-squares framework. This approach can be extended to include symmetry properties, see e.g.\ \cite{mohebujjaman2019,sharma2022a}, and could be a viable alternative to construct energy-stable ROMs without having access to the FOM. 

The system of equations \eqref{eqn:rom_mom_semidiscrete} - \eqref{eqn:rom_energy_semidiscrete} is integrated in time with the same fourth order explicit Runge-Kutta method that was used for the FOM, with the exception that no pressure Poisson equation needs to be solved, since the pressure has disappeared from the ROM formulation.

\subsection{Stability and energy conservation}\label{sec:ROM_stability}
Our proposed ROM formulation \eqref{eqn:rom_mom_semidiscrete}-\eqref{eqn:rom_energy_semidiscrete} has favorable stability characteristics as it mimics the three properties listed for the FOM in Section \ref{sec:FOM_stability}: (i) kinetic-energy conserving pressure gradient, (ii) skew-symmetric momentum convection and (iii) skew-symmetric temperature convection. The first property is trivial to proof because the pressure gradient has disappeared from the formulation. It gives the ROM formulation a strong sense of stability, since any potential issues with inf-sup stability are avoided. 

Secondly, the skew-symmetry of the convective terms in the momentum equations is retained upon POD-Galerkin projection: if $\tilde{C}_{V}(\cdot) = -\tilde{C}_{V}^{T}(\cdot)$, then
\begin{equation}
    \Phi^T \tilde{C}_{V} (\Phi a) \Phi = -\Phi^T \tilde{C}_{V} (\Phi a)^{T} \Phi = - (\Phi^{T} \tilde{C}_{V} (\Phi a) \Phi)^{T},
\end{equation}
which is still skew-symmetric, and the contribution to the kinetic energy equation is 
\begin{equation}
    a^{T} \Phi^T \tilde{C}_{h} (\Phi a) \Phi a = 0,
\end{equation}
just like for the FOM and for the continuous equations.
Thirdly, the skew-symmetry of the convective terms in the temperature equation is also retained upon POD-Galerkin projection: if $\tilde{C}_{\theta} = -\tilde{C}_{\theta}^{T}$, then
\begin{equation}
    \Psi^T \tilde{C}_{\theta} (\Phi a) \Psi = -\Psi^T \tilde{C}_{\theta} (\Phi a)^{T} \Psi = - (\Psi^{T} \tilde{C}_{\theta} (\Phi a) \Psi)^{T},
\end{equation}
which is still skew-symmetric, and its contribution to the equation for the ROM approximation to $\theta^2$ is
\begin{equation}
    b^{T} \Psi^T \tilde{C}_{\theta} (\Phi a) \Psi b = 0,
\end{equation}
just like equation \eqref{eqn:skew_symmetry_theta2}.

Overall, the fact that the skew-symmetric nature of the convective terms is retained by our ROM, combined with the fact that it is pressure-free, makes it suitable for long-time integration, which is necessary in RBC in order to sample statistics. % \FloatBarrier
\section{Results}\label{sec:results}
\begin{figure*}[ht!]
    \centering
    \includegraphics[width=\textwidth]{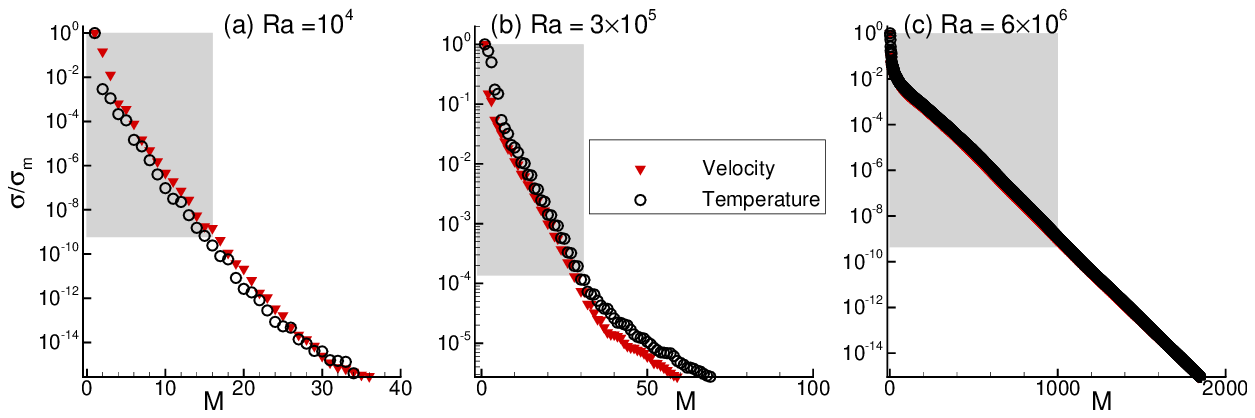}
    \caption{Decay of singular values $(\sigma /\sigma_m)$ with  increasing modes $(M)$ for (a) $\Ray=10^4$, (b) $3\times 10^5$, and (c) $6\times 10^6$. Here, $\sigma_m$ refers to the maximum singular value; the shaded region indicates the maximum number of modes ($M=16,32$ and $1000$ in (a), (b) and (c), respectively) used in the three $\Ray$ cases.}
    \label{fig:energydecay}
\end{figure*}

In this section, we discuss the long-time stability and accuracy of the proposed ROM for the three cases that were listed in Table \ref{tab:results_simdetails}. The reason for studying these cases lies in their different flow dynamics, ranging from steady state (time-independent) to chaotic. For all the FOM cases, we ensure that the grid resolution remains smaller than the Kolmogorov length scale $(\Delta x/\eta< 1)$ and that a sufficient number of grid points constitute the thermal boundary layer $(N_{\text{BL}})$. 
We propose to use the mean $(\langle \theta \rangle)$ and variance $(\sigma_\theta)$ of the temperature as a measure of the ROM accuracy, defined as 
\begin{equation}\label{Eq:mean_variance}
\centering
\langle \theta \rangle (y) = \langle \theta \rangle_{A,t} \hspace{2cm} \text{and} \hspace{2cm} \sigma_\theta (y) = \langle (\theta-\langle \theta \rangle_{t})^2 \rangle_{A,t}
\end{equation}
where $\langle \cdot \rangle_{A,t}$ represents averaging in $x$-direction and in time. These profiles reveal the physical characteristic of the flow. For instance, they both are a measure of thermal boundary layer thickness and characterize the bulk and boundary layer region \cite{Ahlers_2009,Zhou_2011,Chand_2019}. Convergence of the variance, being a second-order statistic, is more difficult than convergence in the mean (a first-order statistic), even for the FOM. Requiring convergence of second-order statistics for the ROM is therefore a good and stringent test to assess its accuracy. 

In addition, we use two global heat transport properties ($\Nu$ and $\Rey$) to investigate the ROM's stability and accuracy. In contrast to temperature mean and variance, these are a still a function of time (but not of space). While $\Nu$ is given by Eq.\ \eqref{Eq:Nu}, $\Rey$ can be defined as
\begin{equation}
    \Rey(t) = \sqrt{\frac{\Ray}{\Pran} \langle \textbf{u}\cdot\textbf{u}\rangle_{V}},
\end{equation}
where $\langle \cdot \rangle_{V}$ refers to volume averaging. These two quantities are the primary objective of any RBC study. 

Before starting the discussion of the results, we first describe the decay of the singular values in the three cases, see Fig. \ref{fig:energydecay}. The singular values (with respect to weighted inner product) represent the kinetic energy of the system. The singular value decay helps in the selection of the number of modes required to approximate the FOM, where the selection process is solely based on capturing the total kinetic energy. It is apparent that $M=16$, $32$, and \textcolor{red}{$200$} modes are sufficient to capture at least $99\%$ of the kinetic energy in $\Ray=10^4$, $3\times10^5$ and $6\times 10^6$, respectively, see the shaded region. 
The singular values are computed by performing an SVD of the snapshot matrix. It is important to mention that the snapshot matrix is constructed from the start of the simulation $(t=0)$ for the steady state case, whereas, for the other two cases, it is constructed after attaining the statistical steady state, see $T_{\text{sampling}}$ in Table \ref{tab:results_simdetails}. This state begins once the flow passes the transient period, and it is identified by stable $\Nu$ and $\Rey$ time series. 

Note that in the results presented here we are not reporting the computational speed-up achieved by the ROM compared to the FOM, as we focus on the stability and accuracy properties of the method. However, as the ROM proposed here is an extension of the ROM proposed in \cite{Benjamin_2020}, we can expect very similar speed-ups. For example, with $M=8$ the speed-up in that work was $\mathcal{O}(10^{3})$.

\subsection{Steady flow case}\label{case1}

\begin{figure}[ht]
\centering
     \begin{subfigure}[t]{0.4\textwidth}
\includegraphics[width=\textwidth]{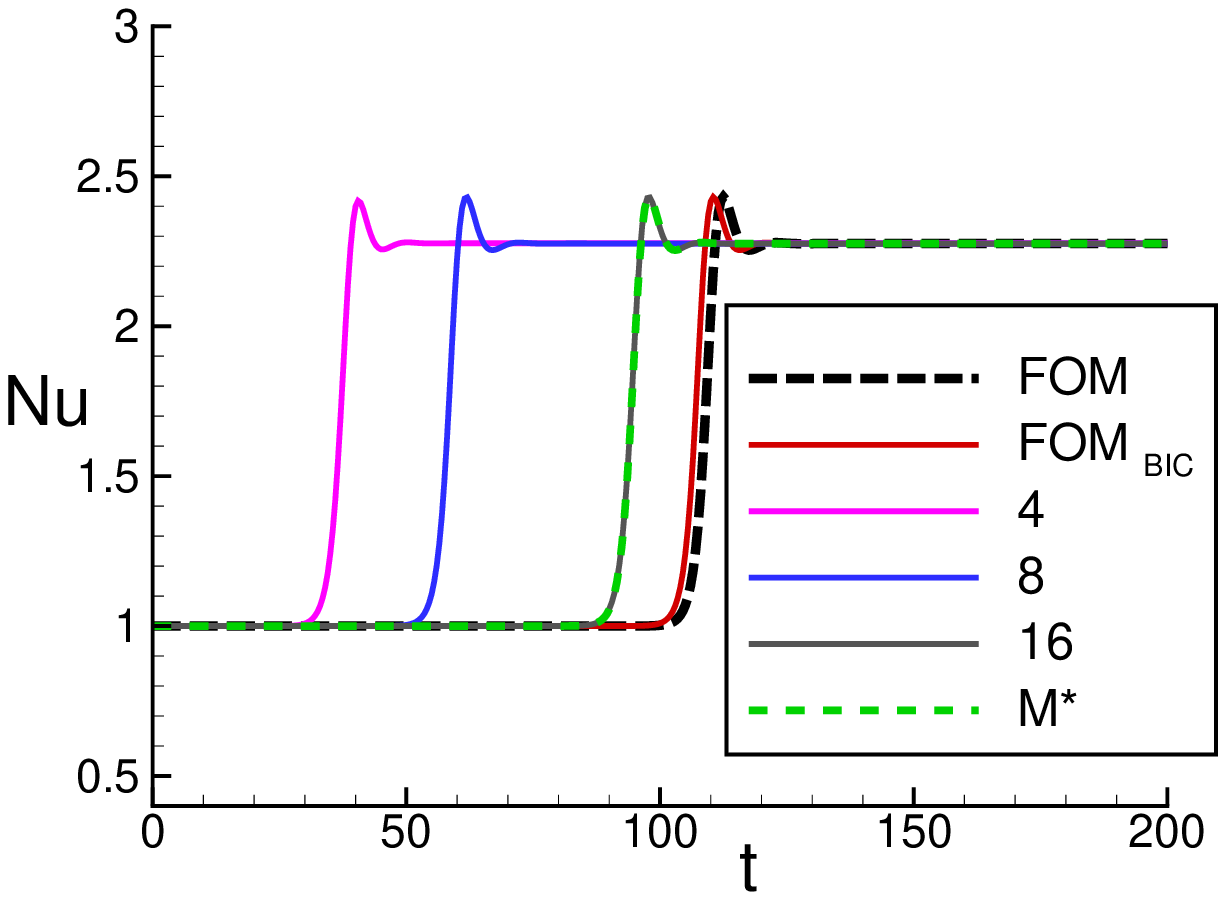}
\caption{$\Nu$ time series.}
\label{fig:Nu_steady}
 \end{subfigure}
 \hfill
  \begin{subfigure}[t]{0.4\textwidth}
\includegraphics[width=\textwidth]{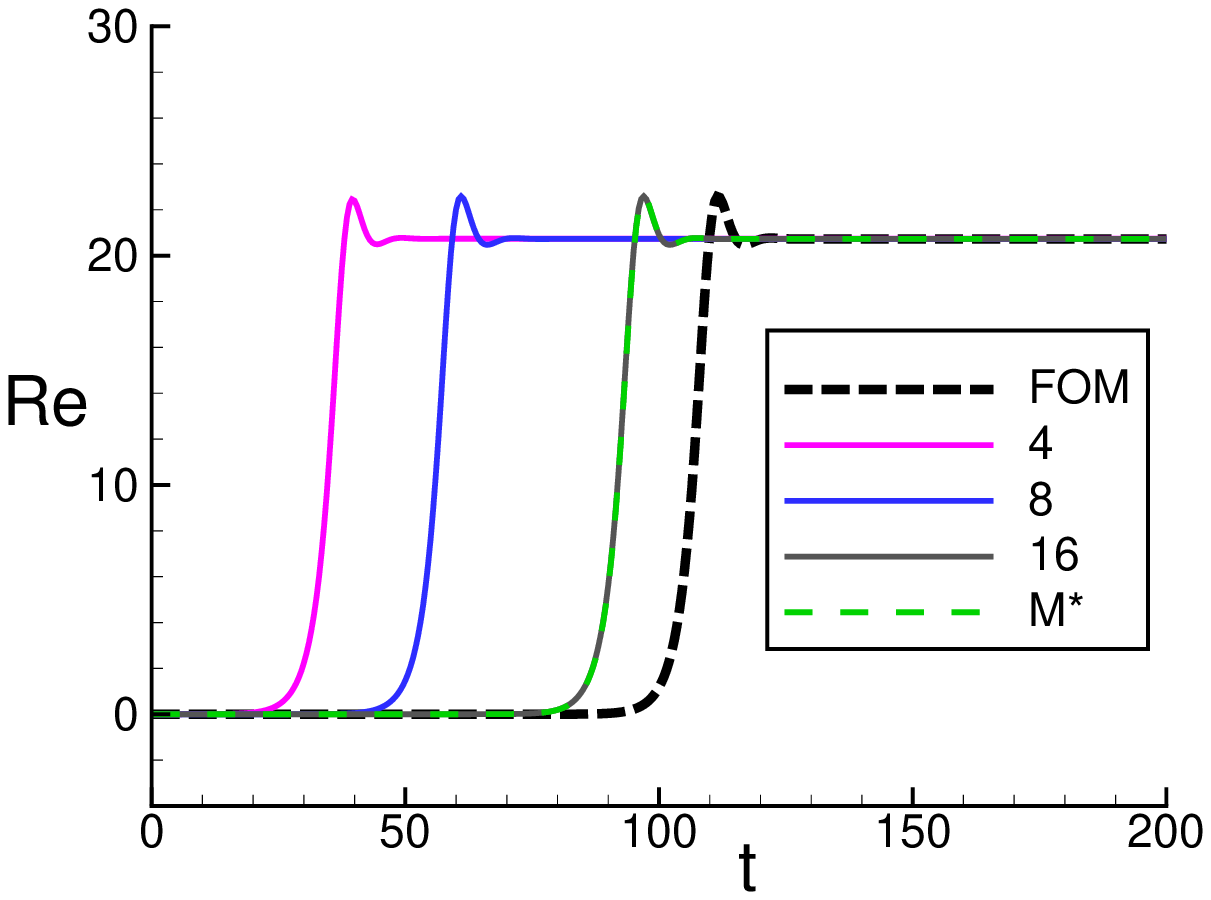}
\caption{$\Rey$ time series.}
\label{fig:Re_steady}
 \end{subfigure}
\caption{Effect of initial condition and $M$ on the onset of convection at $\Ray=10^{4}$. Here FOM$_\text{BIC}$ refers to the FOM case whose initial condition is the best approximation computed by considering all the modes $(M^*=4560)$.}% and given by Eqs. \ref{Eq:error_split}-\ref{Eq:bestapprox}.}
\end{figure}

We start the discussion by studying the time series parameters $\Nu$ and $\Rey$. Figure \ref{fig:Nu_steady} shows the $\Nu$ time series for different cases: (i) the FOM, (ii) the FOM with initial condition as the best approximation solution (FOM$_{\text{BIC}}$), and (iii) the ROM with different number of modes. Figure \ref{fig:Nu_steady} shows that the evolution of $Nu$ time series is different in all the cases. Figure \ref{fig:Re_steady} shows a similar result, but displaying $\Rey$ instead of $\Nu$. 
The difference between the cases seems remarkable, given that the difference between the exact initial condition and its best approximation is of order $10^{-14}$. This minimal difference in the initial condition triggers a slightly early onset to convection. However, when realising that the instability arises from the growth of round-off errors, it is quite logical that a perturbation in machine precision changes the onset of instability. This difference could be avoided by triggering a certain mode in the initial condition, for example the most unstable mode that follows from linear stability analysis. This is not done here on purpose, as we aim to show the chaotic nature of the test case and the importance of horizontal and temporal averaging. 

A similar behavior can be noted for the ROM cases: the fewer the number of modes, the earlier is the onset to convection. This indicates that simulations that are chaotic (highly sensitive to initial conditions or parameter values) can yield rather different results even for a very small perturbation, as caused for example by the approximation of the FOM by the ROM. Consequently, in order to assess the accuracy of the ROM it is important to use error metrics that involve averaging in time, like equation \eqref{Eq:mean_variance}. 

\begin{figure}[ht]
\centering
\includegraphics[width=0.5\textwidth]{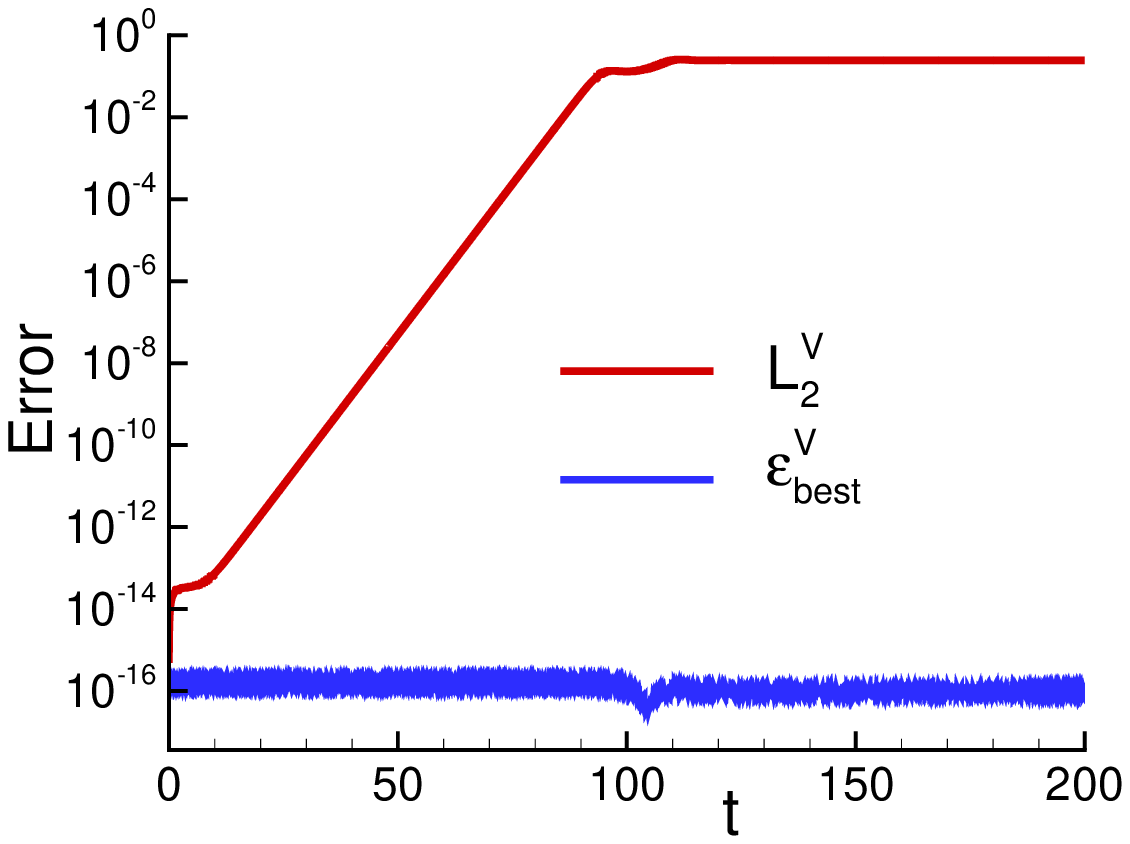}
\caption{Comparison between evolution of error in ROM velocity with all the modes $M^*=4560$ $(L_2^V)$ and error obtained by projecting FOM velocity field onto the POD modes $(\epsilon_\text{best}^V)$, see text for their definition.}
\label{fig:L2error}
\end{figure}

In order to further clarify this point, we split the error in the ROM solution that employs all modes ($M=M^{*}$) as
\begin{equation}\label{Eq:error_split}
    V_{r}^{n} - V_{h}^{n} = \underbrace{V_{r}^{n} - V_{\text{best}}^{n}}_{\textrm{temporal error}} +  \underbrace{V_{\text{best}}^{n} - V_{h}^{n}}_{\textrm{basis error due to SVD precision}}, 
\end{equation}
where $V_\text{best}^n = \Phi \Phi^T \Omega_h V_h^n$. We define the errors 
\begin{equation}\label{Eq:bestapprox}
\centering
L_2^V = \frac{\|V_r^n-V_h^n\|_{\Omega_h}}{\|V_\text{ref}\|_{\Omega_h}} \hspace{1cm} \text{and} \hspace{1cm} \epsilon^V_\text{best} = \frac{\|V_\text{best}^n-V_h^n\|_{\Omega_h}}{\|V_\text{ref}\|_{\Omega_h}},
\end{equation}
where $V_\text{ref}$ and $\| \cdot \|_{\Omega_h}$ represent a reference velocity and the $\Omega_h$ norm, respectively. Figure \ref{fig:L2error} shows a comparison of the evolution of the error in the ROM velocity field $(L_2^V)$ when all modes are included ($4560$ in total), and the best approximation $(\epsilon^V_\text{best})$. The best approximation error is at machine precision. The $L_2^V$ error also starts with a magnitude of order of machine precision but grows significantly (owing to a horizontal shift in the large-scale structure due to early onset of convection). In view of equation \eqref{Eq:error_split}, this means that the temporal error $V_{r}^{n} - V_{\text{best}}^{n}$ is dominating the evolution of $L_{2}^{V}$. Therefore, comparing the $L_2^{V}$ and $\epsilon_\text{best}^V$ errors does not give a good measure of the accuracy of the ROM. This is in contrast to what was concluded for a non-chaotic system in \cite{Benjamin_2020}. 

\begin{figure*}[h]
\centering
\includegraphics[width=\textwidth]{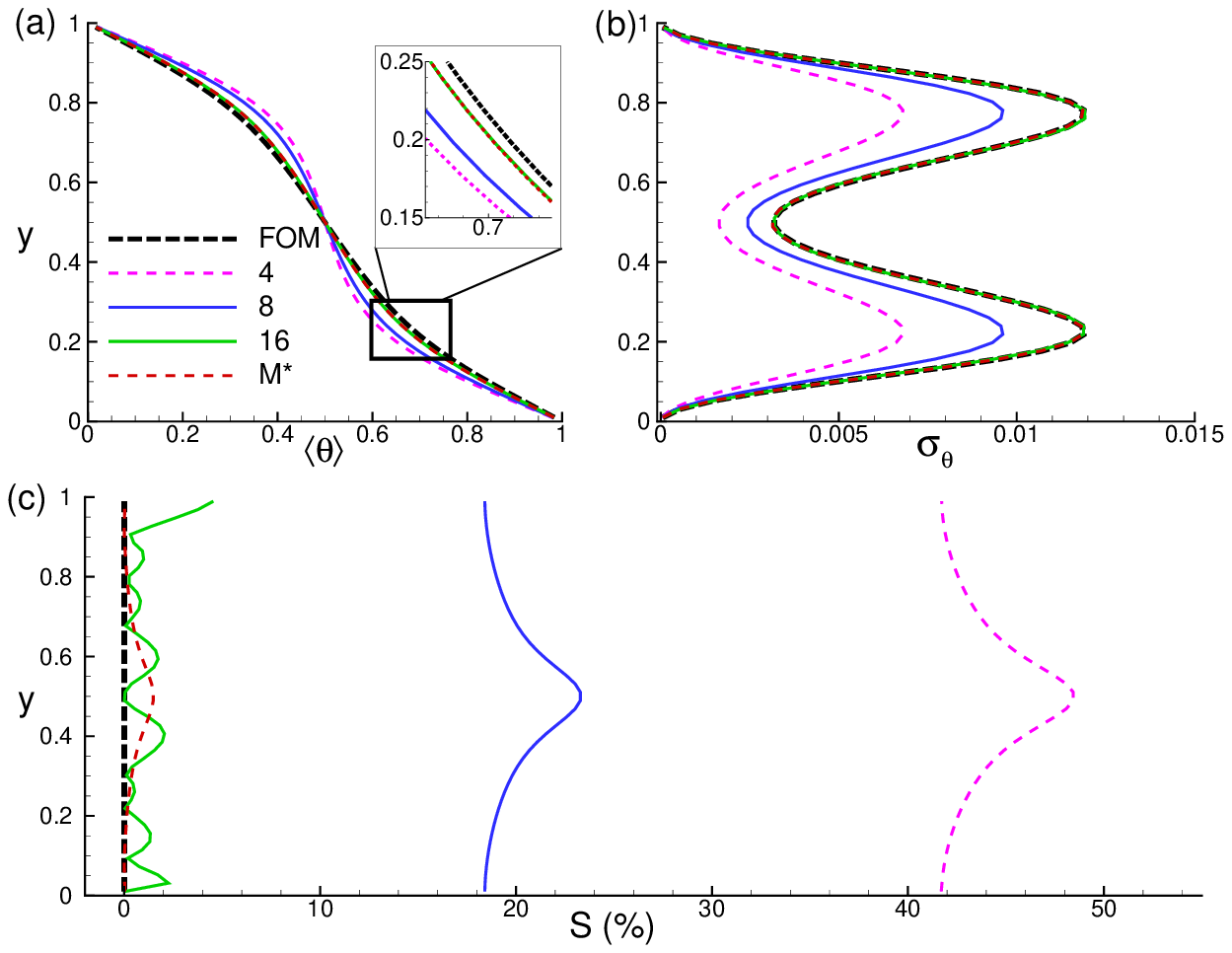}
\caption{For $\Ray=10^4$, vertical profiles of (a) mean temperature $\langle \theta \rangle$, (b) variance of temperature $\sigma_\theta$, and (c) relative error $(S)$ in the ROM profiles of $\sigma_\theta$ with respect to the FOM. Inset shows the zoomed-view.}
\label{fig:meanvar_1e4}
\end{figure*}

Instead, to assess the ROM accuracy in chaotic flows, we propose to employ statistical quantities, namely the first two moments (mean and variance) of the temperature as defined in equation \eqref{Eq:mean_variance}. 
In Fig. \ref{fig:meanvar_1e4}, the effect of $M$ on the vertical profiles of mean and variance of temperature is shown. In comparison to the $\Nu$ and $\Rey$ time series, clearer convergence is obtained in the vertical profiles of mean temperature, where the ROM profiles shift towards that of the FOM as $M$ increases. Furthermore, the vertical profiles of the variance indicate that the temperature statistics obtained with the ROM also converges upon increasing $M$. The profiles nearly overlap with the FOM data when all the modes are considered. To elucidate this, in Fig. \ref{fig:meanvar_1e4}, we quantify the error in the ROM profiles as
\begin{equation}
\centering
S=\frac{|\sigma_{\theta_{\text{FOM}}}-\sigma_{\theta_{\text{ROM}}}|}{|\sigma_{\theta_{\text{FOM}}}|} \times 100.
\end{equation}
These profiles clearly show diminishing errors with increasing number of modes. In particular, for $M\geq16$, the error drops below $2\%$. Only a slight difference close to the top wall is observed. 

Note that the relatively large errors for $M=4$ and $M=8$ are mainly due to the difference in onset time (see figure \ref{fig:Nu_steady}). As was shown, this difference in onset time persists even for $M=M^{*}$, due to the chaotic nature of the system, and consequently a small error remains in the vertical profiles even for $M=M^{*}$. If one would time-average over only the last few time units, then the profiles would be very close, as only the best approximation error would remain.

\begin{figure*}[h!]
\centering
     \begin{subfigure}[b]{0.45\textwidth}
         \centering
\includegraphics[width=\textwidth]{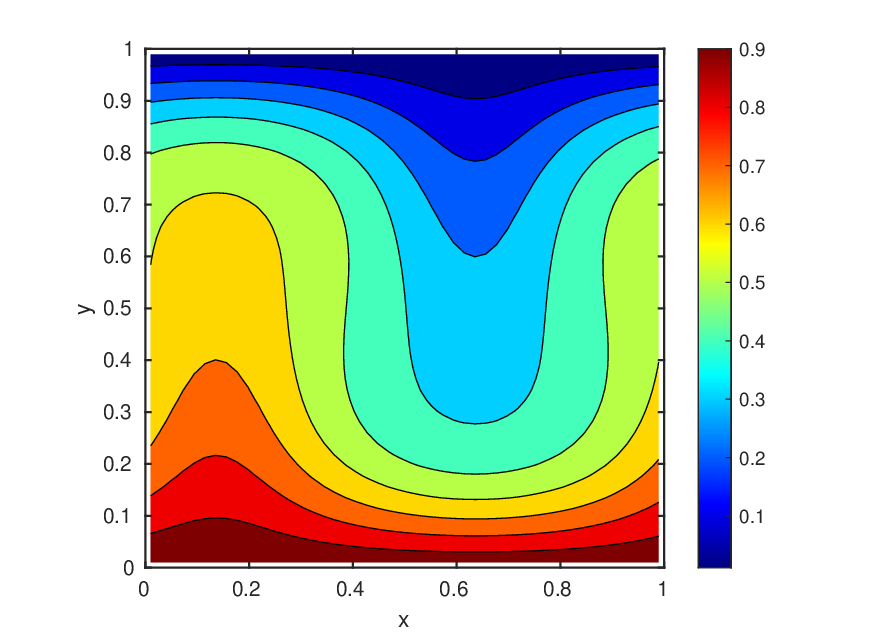}
\caption{FOM}
     \end{subfigure}
     \hfill    
     \begin{subfigure}[b]{0.45\textwidth}
         \centering
\includegraphics[width=\textwidth]{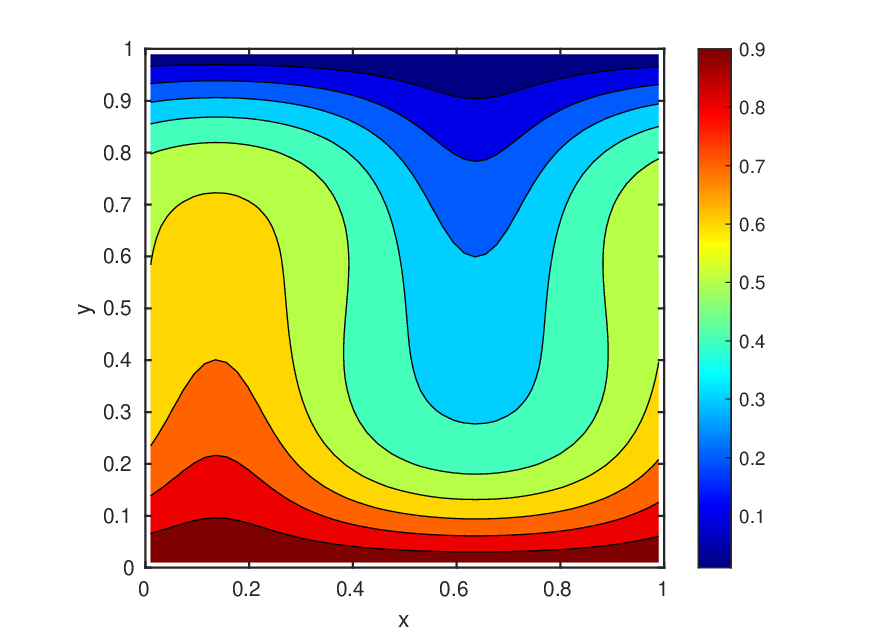}
\caption{ROM with $M= 4$}
     \end{subfigure}
     \hfill
     \begin{subfigure}[b]{0.45\textwidth}
         \centering
\includegraphics[width=\textwidth]{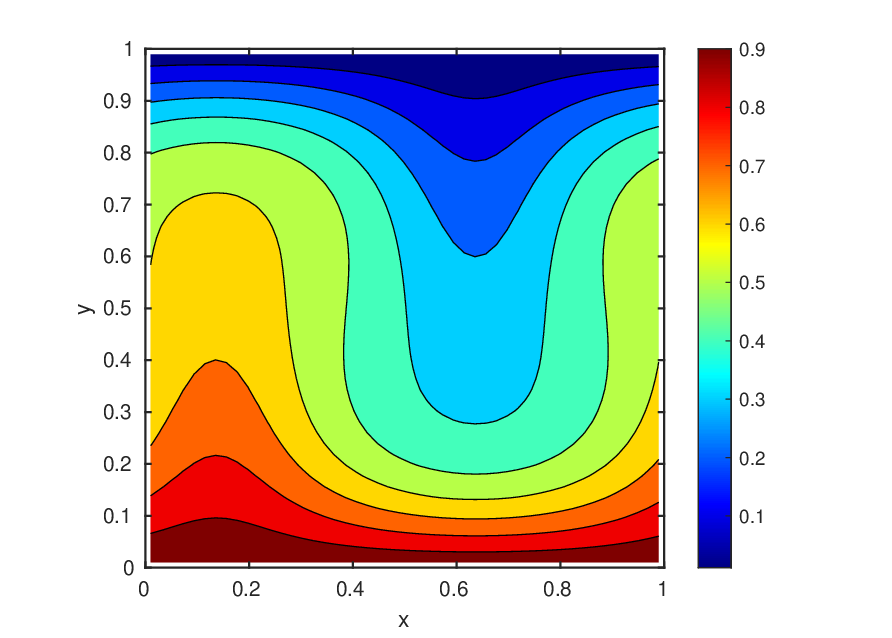}
\caption{ROM with $M= 8$}
     \end{subfigure}
     \hfill
     \begin{subfigure}[b]{0.45\textwidth}
         \centering
\includegraphics[width=\textwidth]{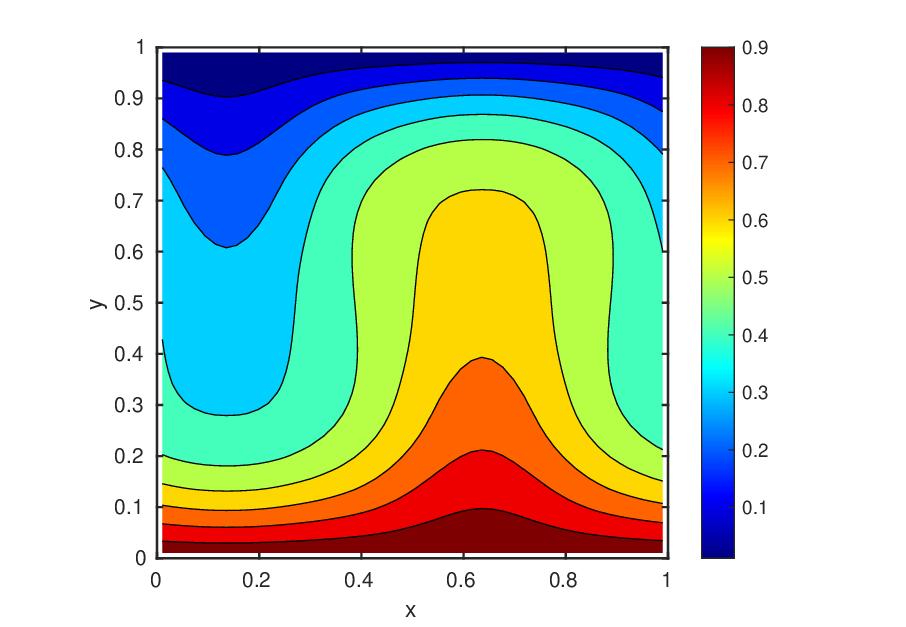}
\caption{ROM with $M= 16$}
     \end{subfigure}
     \hfill
     \begin{subfigure}[b]{0.45\textwidth}
         \centering
\includegraphics[width=\textwidth]{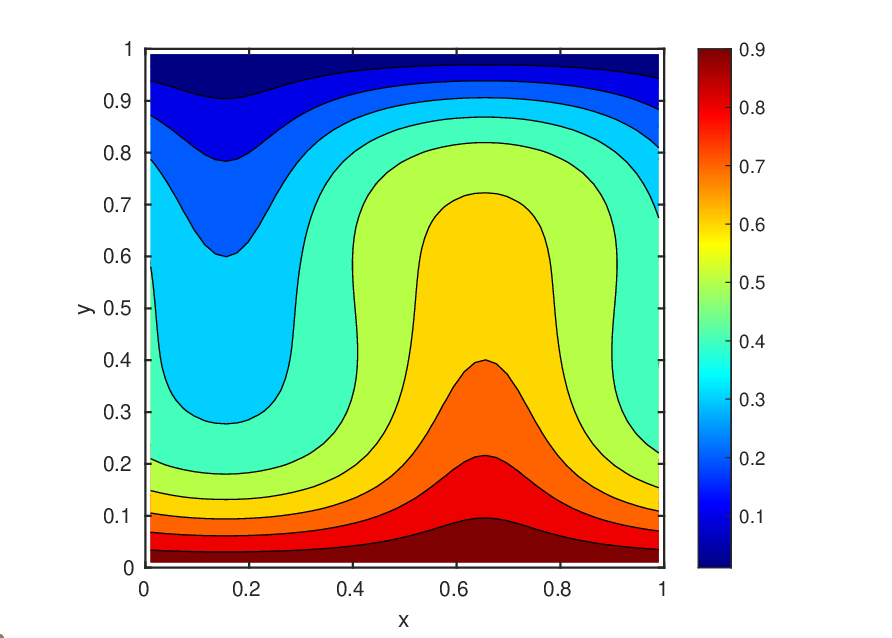}
\caption{ROM with all the modes $(M=M^*)$}
     \end{subfigure}     
     \hfill
\caption{Comparison of instantaneous temperature field of the last time instance for $\Ray=10^4$.} 
\label{fig:meantempfield_1e4}
\end{figure*}

Lastly, we show why it is important to perform horizontal averaging. Fig. \ref{fig:meantempfield_1e4} shows the mean temperature field for different numbers of modes. As expected, flow features appear nearly similar in all the cases because they capture more than $99\%$ of kinetic energy. However, a horizontal offset in the field is evident which can be attributed to a slightly different onset of convection in combination with periodic boundary conditions in lateral direction. Upon horizontal averaging, such effects do not influence the error computation negatively.

% \FloatBarrier

\subsection{Periodic flow case}\label{case2}
\begin{figure*}[ht!]
\centering
\includegraphics[width=\textwidth]{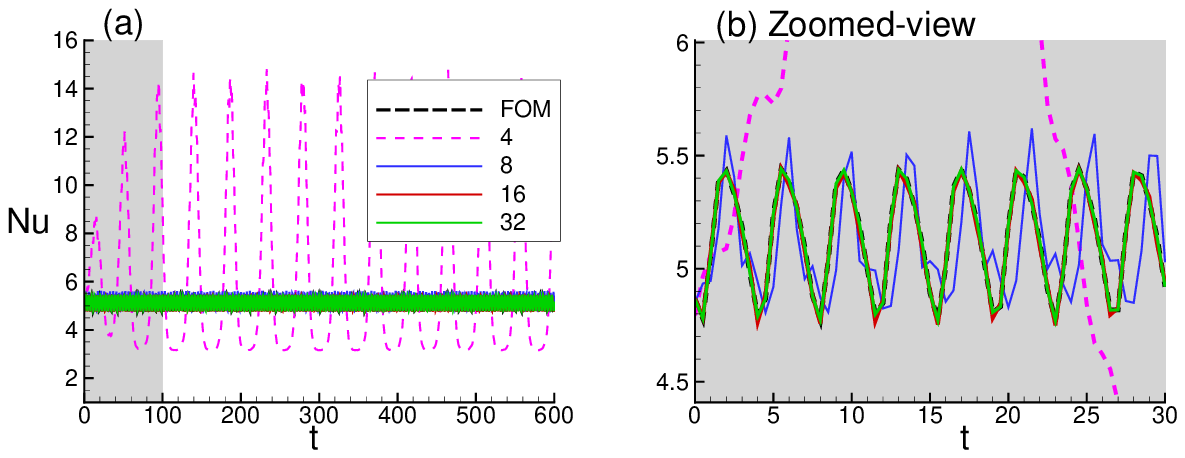}
\caption{(a) $\Nu$ time series for $\Ray=3\times10^5$. (b) Zoomed-view of the time series shown in (a). The grey shaded region indicates the data used to compute POD basis from FOM. Note that the time step for both ROM and FOM is the same (0.01), but the solution is only plotted every 50th timestep.} 
\label{fig:Nu_3e5}
% \end{figure}
% \begin{figure}
% \centering
\includegraphics[width=\textwidth]{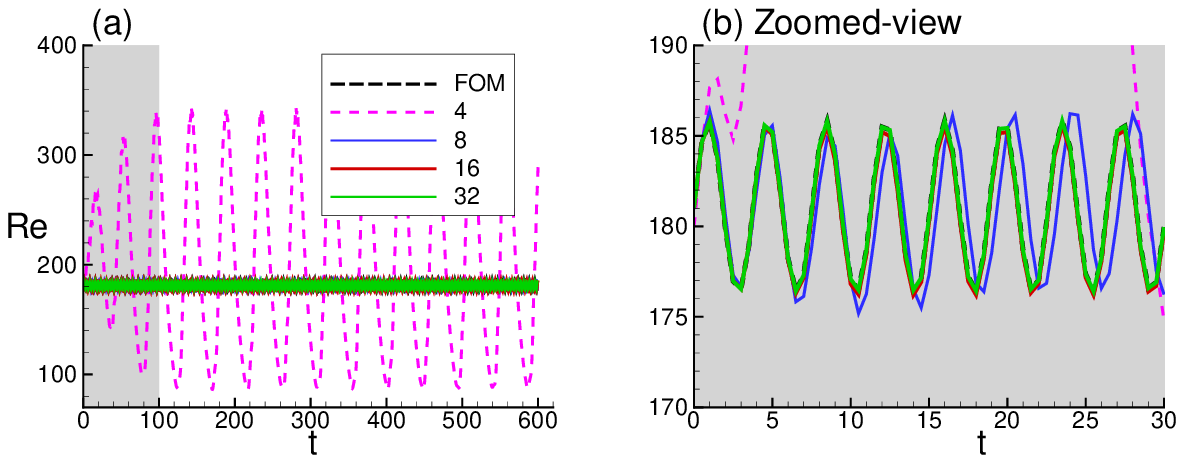}
\caption{(a) $\Rey$ time series for $\Ray=3\times10^5$. (b) Zoomed-view of the time series shown in (a). The grey shaded region indicates the data used to compute POD basis from FOM. Note that the time step for both ROM and FOM is the same (0.01), but the solution is only plotted every 50th timestep.} 
\label{fig:Re_3e5}
\end{figure*}

\begin{table*}[ht!]
\centering
\begin{tabular}{*7c}
\toprule
& \multicolumn{3}{c}{$\Nu$} & \multicolumn{3}{c}{$\Rey$}\\ \cmidrule(lr){2-4}  \cmidrule(lr){5-7}
$M$ & Mean $(\langle \Nu \rangle_t)$ &  $\Nu_\text{error} (\%)$  & Amplitude (range) & Mean $(\langle \Rey \rangle_t)$&  $\Rey_\text{error}(\%)$ &  Amplitude (range) \\
\midrule
FOM  &  $5.17$& $0.00$ & $4.75-5.5$  & $180.64$& $0.00$ & $175.5-185.5$ \\
$4$  &  $9.47$& $83.17$& $3.19-15.74$& $220.05$& $21.82$& $81.04-359.05$ \\
$8$  &  $5.00$& $3.29$ & $4.85-5.45$ & $180.81$& $0.09$ & $175.5-186.5$ \\
$16$ &  $5.18$& $0.19$ & $4.75-5.5$  & $180.37$& $0.15$ & $175.5-186.5$ \\
$32$ &  $5.17$& $0.00$ & $4.75-5.5$  & $180.63$& $0.01$ & $175.5-186.5$ \\
\bottomrule
\end{tabular}
\caption{$\Nu$ and $\Rey$ comparison for $\Ray=3\times10^5$.  From left to right: Modes; time average heat flux $\langle \Nu \rangle_{t}$; error in $\langle Nu \rangle_t$ $(\Nu_\text{error}=|(1-\langle\Nu_\text{ROM}\rangle_t/\langle\Nu_\text{FOM}\rangle_t)|\times100)$; range of $\Nu$; time average Reynolds number $\langle \Rey \rangle_t$; error in $\langle\Rey\rangle_t$ $(\Rey_\text{error}=|(1-\langle\Rey_\text{ROM}\rangle_t/\langle\Rey_\text{FOM}\rangle_t)|\times100)$; and range of $\Rey$.}
\label{tab:range_NuRe_3e5}
\end{table*}

Next, we discuss the periodic flow case at $\Ray=3\times10^5$. We first investigate again the convergence of the ROM through $\Nu$ and $\Rey$ time series and then through the statistics of the temperature. Note that $100$ time units of the FOM are used to build the snapshot matrix (and corresponding basis), which is subsequently used to run the ROM for $600$ time units (effectively extrapolating in time). Figure \ref{fig:Nu_3e5} shows the $\Nu$ time series. All ROM cases yield a periodic $\Nu$ time series, see zoomed-view (Frame b) for higher modes ($M>4$). It is observed that the $\Nu$ series with $M=4$ deviates significantly from the FOM, which can be attributed to the lower amount of energy captured by these few modes, see Fig. \ref{fig:energydecay}b. Compared to the previous test case ($\Ray=10^{4}$), this can be attributed to the fact that the singular value decay is much slower. Importantly, even with such a low number of modes, the solution stays stable when marching in time, without requiring any stabilization mechanism. When the ROM simulation is still inside the training interval $[0,100]$, we see clear convergence towards the FOM solution upon increasing the number of modes. Outside the training interval, point-wise convergence in time is lost, but the amplitude of the oscillations of the ROM still converges to that of the FOM upon increasing the number of modes. To elucidate this, we show the range of the amplitude of oscillations and the corresponding mean values for both $\Nu$ and $\Rey$ in Table \ref{tab:range_NuRe_3e5}. It is evident that with increasing $M$, the mean value approaches the value of the FOM. Similar results are reflected from the $\Rey$ time series, see Fig. \ref{fig:Re_3e5}. The mean $\Rey$ for the FOM and the ROM with $M=32$ are nearly equal, and the range of amplitude is also similar beyond $M=8$. The error drops below $1\%$ beyond $M=8$.

\begin{figure*}[ht!]
\centering
\includegraphics[width=0.9\textwidth]{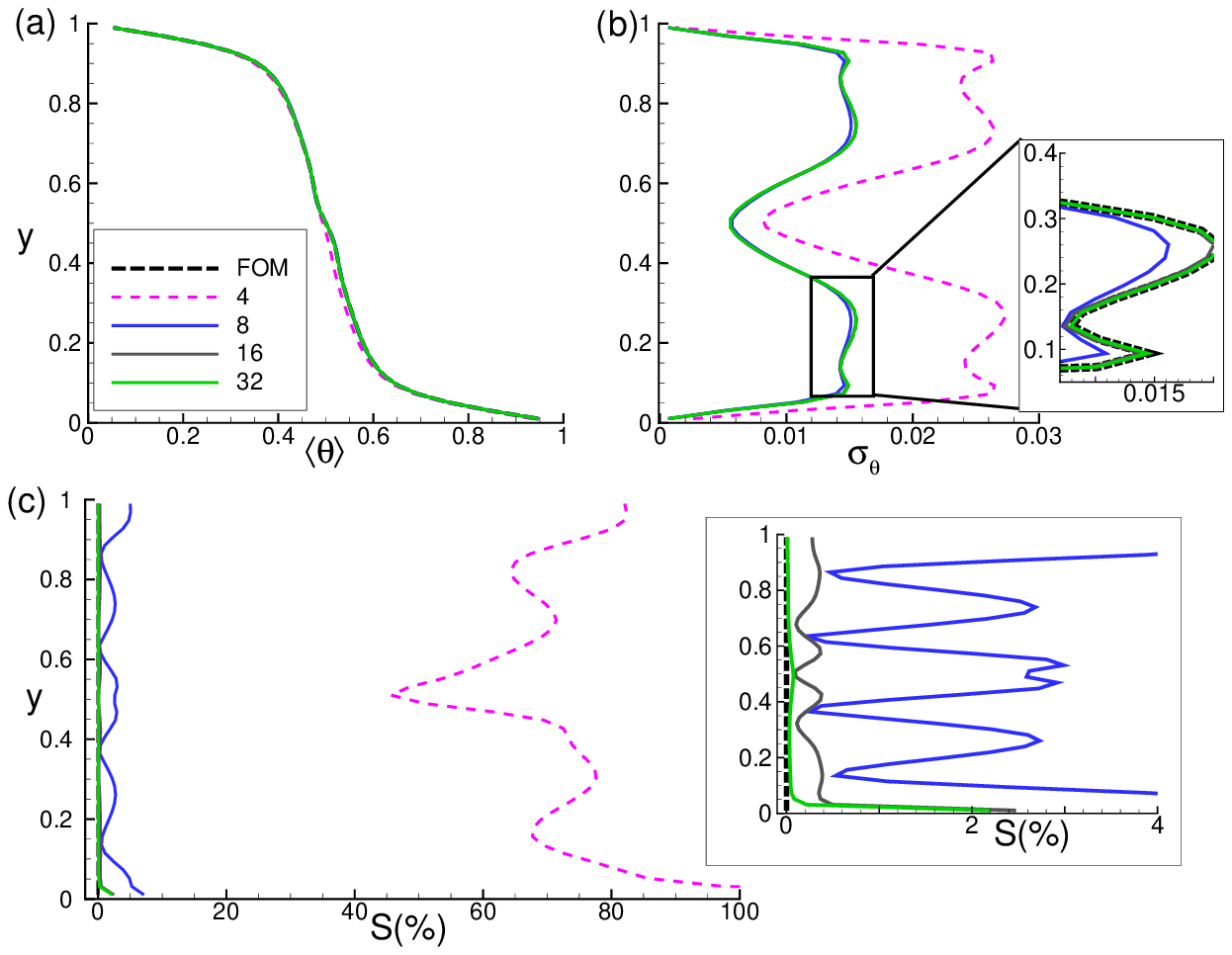}
\caption{For $\Ray=3\times10^5$, vertical profiles of (a) mean temperature $\langle \theta \rangle$, (b) variance of temperature $\sigma_\theta$, and (c) relative error $(S)$ in the ROM profiles of $\sigma_\theta$ with respect to the FOM. Inset shows the zoomed-view.}
\label{fig:meanvar_3e5}
\end{figure*}

Figure \ref{fig:meanvar_3e5} compares the effect of $M$ on the statistics of the temperature. Roughly, the vertical profiles of $\langle \theta \rangle$ of all the ROM cases (except for $M=4$) overlap with those of the FOM. Moreover, the vertical profiles of the variance show a clear convergence, see the inset in Fig. \ref{fig:meanvar_3e5}b, where the ROM profiles shift towards that of the FOM as $M$ increases. The convergence is further confirmed by a diminishing relative error $S$, which is less than $5\%$ beyond $M=4$ in almost the entire domain. In comparison with the previous (steady flow) test case, we observe again that the main dynamics and statistics are correctly captured even though the time series of the FOM are not exactly reproduced in a point-wise sense. 

\begin{figure*}[ht!]
\centering
     \begin{subfigure}[b]{0.45\textwidth}
         \centering
\includegraphics[width=\textwidth]{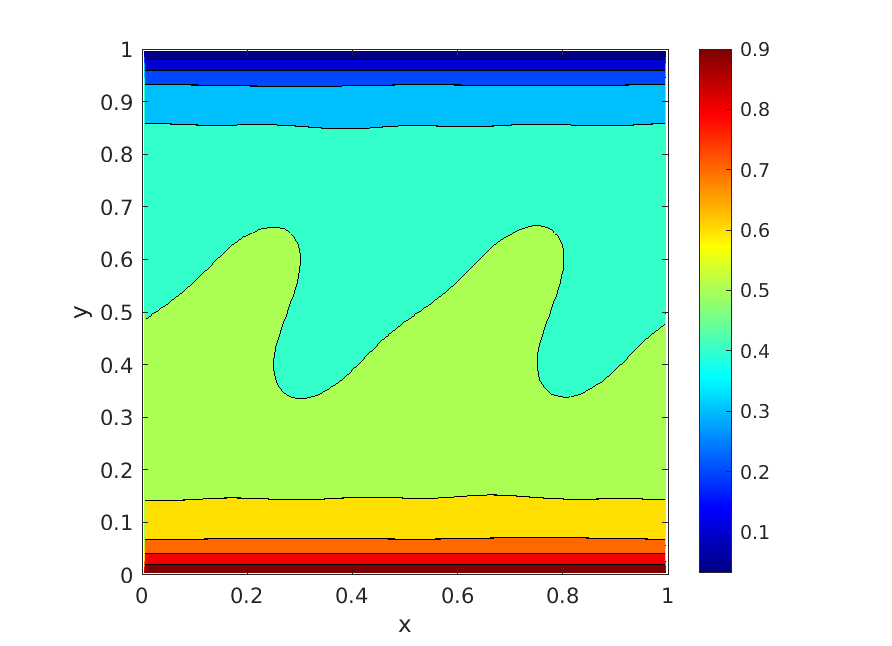}
\caption{FOM}
     \end{subfigure}
     \hfill
     \begin{subfigure}[b]{0.45\textwidth}
         \centering
\includegraphics[width=\textwidth]{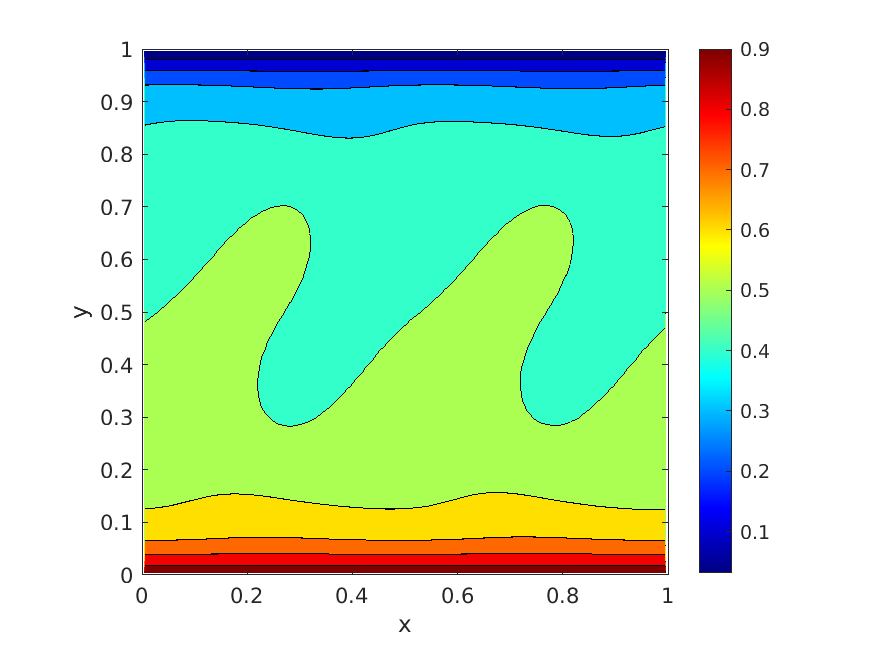}
\caption{ROM with $M= 4$}
     \end{subfigure}
     \hfill
     \begin{subfigure}[b]{0.45\textwidth}
         \centering
\includegraphics[width=\textwidth]{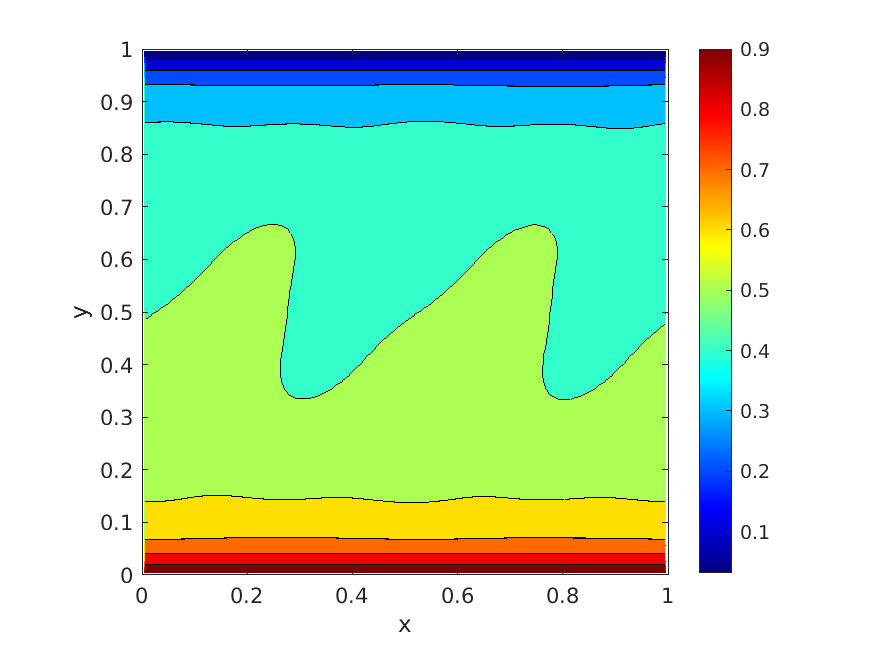}
\caption{ROM with $M= 8$}
     \end{subfigure}
     \hfill
     \begin{subfigure}[b]{0.45\textwidth}
         \centering
\includegraphics[width=\textwidth]{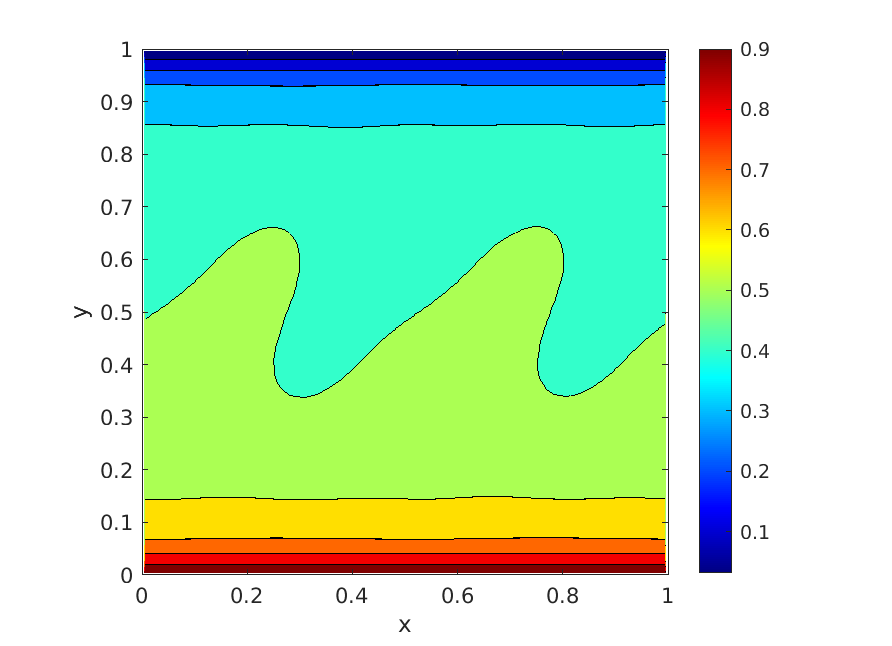}
\caption{ROM with $M= 16$}
     \end{subfigure}
     \hfill
     \begin{subfigure}[b]{0.45\textwidth}
         \centering
\includegraphics[width=\textwidth]{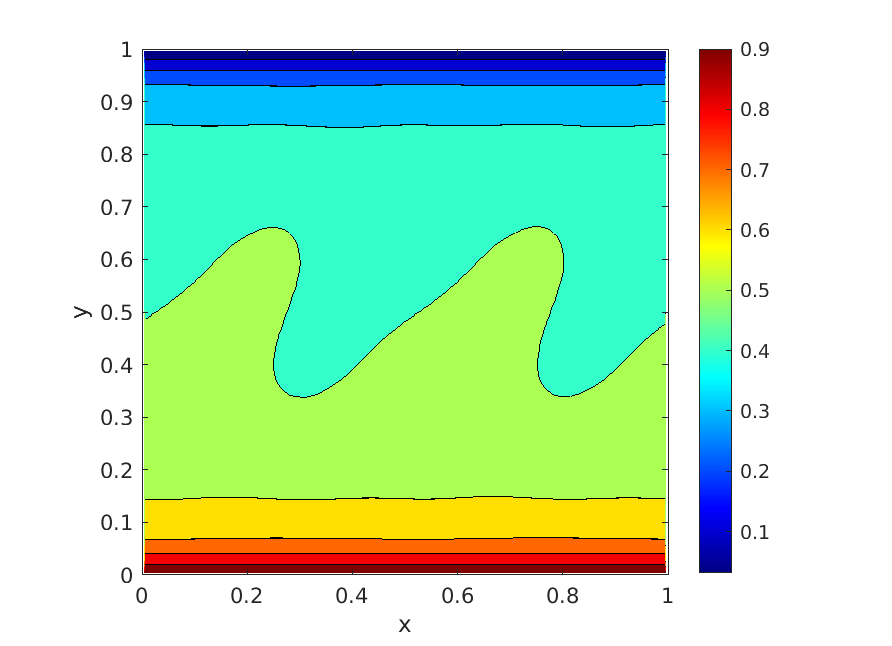}
\caption{ROM with $M= 32$}
     \end{subfigure}
     \hfill
\caption{Comparison of mean temperature field for $\Ray=3\times10^5$. The averaging is carried out over $550$ time units.} 
\label{fig:meantempfield_3e5}
\end{figure*}

Figure \ref{fig:meantempfield_3e5} shows the mean temperature field for the FOM and the ROM cases. A periodic finger-like structure is evident in all the cases. In the smallest mode case, the structures appear to be more elongated as compared to other higher modes and FOM cases. Moreover, the near-wall regions seem relatively more disturbed. These variations can be attributed to the fact that insufficient kinetic energy is captured by the ROM with $M=4$ (see Fig. \ref{fig:energydecay}). Fortunately, when more modes are included, the flow structures in the ROM closely resemble those of the FOM, and already at $M=16$ the mean field looks visually indistinguishable. Note that the horizontal shift that was observed in Fig. \ref{fig:meantempfield_1e4} is not present in this simulation.

% \FloatBarrier

\subsection{Chaotic flow case}\label{case3}
\begin{figure*}[hbtp]
\centering
\includegraphics[width=0.9\textwidth]{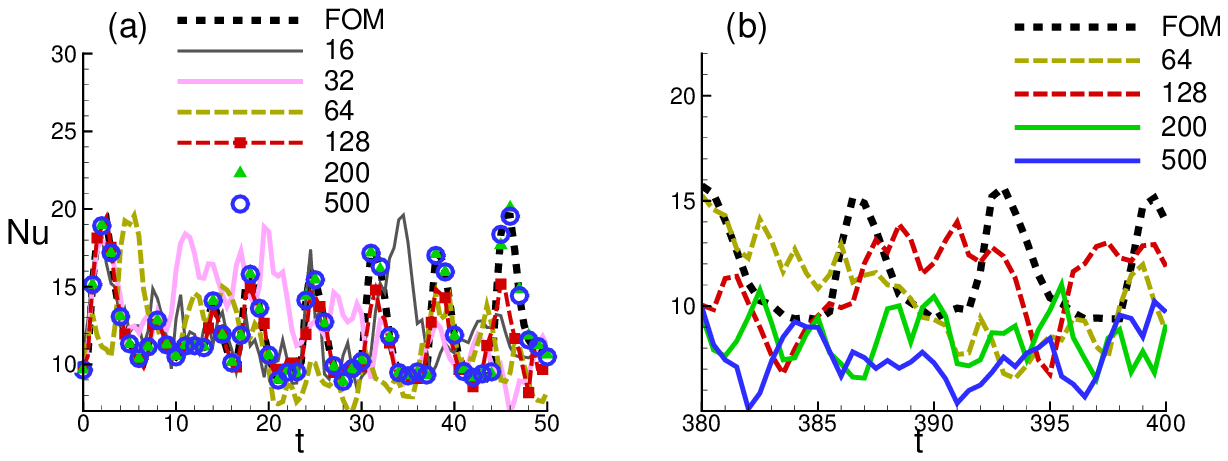}
\caption{(a) $\Nu$ time series for $\Ray=6\times10^6$ for the training interval; (b) same, outside the training interval, indicating long-time stable time series.}
\label{fig:Nu_6e6}
\end{figure*}

\begin{figure*}
\centering
\includegraphics[width=0.9\textwidth]{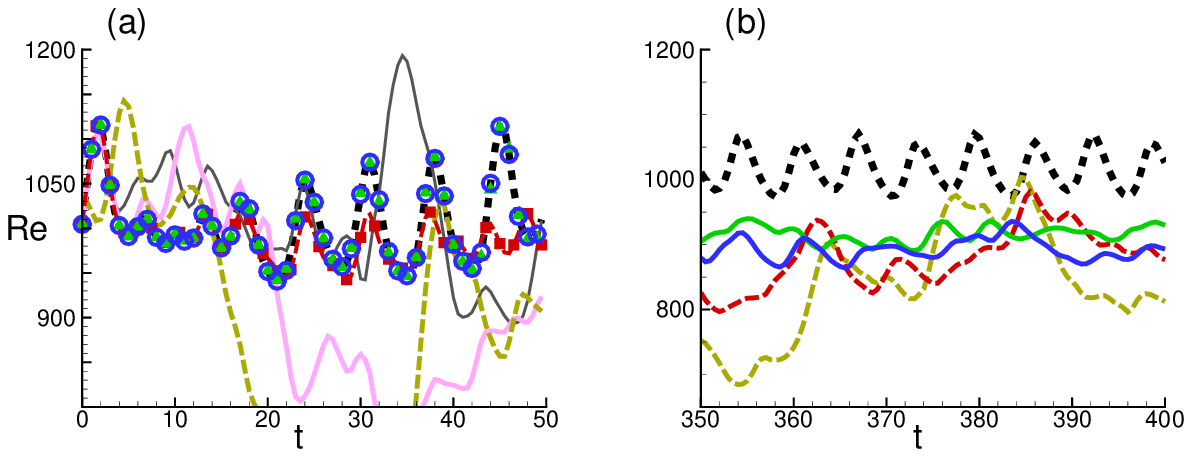}
\caption{(a) $\Rey$ time series for $\Ray=6\times10^6$ for the training interval; (b) same, outside the training interval, indicating long-time stable time series. For symbols, refer Fig. \ref{fig:Nu_6e6}.}
\label{fig:Re_6e6}
\end{figure*}

We finally investigate the convergence of the ROM for a chaotic case $(\Ray=6\times 10^6)$. This case typifies turbulent flow characteristics and we again discuss accuracy and stability of the ROM. Figures \ref{fig:Nu_6e6} and \ref{fig:Re_6e6} show the evolution of $\Nu$ and $\Rey$, respectively. Contrary to the previous case, neither the $\Nu$ nor the $\Rey$ time series obtained with the FOM show clear periodic characteristics, which justifies the reason for selecting this case. 

Our first observation is that our ROM yields a stable solution and corresponding time series for $\Nu $ and $\Rey$, independent of the number of modes chosen. Note that only 50 time units were used for constructing the snapshot matrix, whereas the ROM is simulated much longer in time, namely until 400 time units. Also in this extrapolation scenario, our proposed ROM yields stable results.  


In contrast to the periodic flow case, a qualitative inspection shows that the time series do \textit{not} indicate a clear mode convergence at smaller modes. However, for higher mode cases (greater than 128), both the time series of ROM collapse onto that of FOM (within the training period). In order to make this more quantitative, we compute the mean global transport properties and their relative error, see Table \ref{tab:NuRe_6e6}. The relative error does not show a consistent mode convergence trend up to $M=64$, however, beyond it, clear convergence is observed and  the error drops to $0.00\%$ for $M=500$ in both $\Nu$ and $\Rey$. This shows that for chaotic cases, a small number of modes is not sufficient to approximate the flow dynamics due to discarding the lower energy structures in dissipating systems like turbulent flows. 
Here, it is important to emphasise the long-time stable $\Nu$ and $\Rey$ time-series, which goes far beyond the training data set (up to $400$ time units).

\begin{table}[hbtp]
\centering
\begin{tabular}{*5c}
\toprule
& \multicolumn{2}{c}{$\Nu$} & \multicolumn{2}{c}{$\Rey$}\\ \cmidrule(lr){2-3}  \cmidrule(lr){4-5}
Modes $(M)$& $\langle \Nu 
\rangle_{t}$ & $\Nu_\text{error} (\%)$ & $\langle\Rey\rangle_{t}$ & $\Rey_\text{error}(\%)$ \\
\midrule
FOM   & $12.27$ & $0.00	 $ & $1006.32$ & $0.00 $\\
$4$   & $26.53$ & $116.22$ & $1656.20$ & $64.58$\\ 
$8$   & $15.68$ & $27.79 $ & $1112.06$ & $10.51$\\
$16$  & $12.25$ & $0.16	 $ & $1013.87$ & $0.75 $\\
$32$  & $12.84$ & $4.65	 $ & $922.55 $ & $8.32 $\\
$64$  & $11.14$ & $9.21	 $ & $887.73 $ & $11.78$\\
$128$ & $11.71$ & $4.56	 $ & $991.812$ & $1.44 $\\
$200$ & $12.28$ & $0.08	 $ & $1006.03$ & $0.03 $\\
$500$ & $12.27$ & $0.00	 $ & $1006.32$ & $0.00 $\\
\bottomrule
\end{tabular}
\caption{$\Nu$ and $\Rey$ comparison for $\Ray=6\times10^6$ for the training data. From left to right: Modes; time average heat flux $\langle \Nu \rangle_{t}$; error in $\langle Nu \rangle_t$ $(\Nu_\text{error}=|(1-\langle\Nu_\text{ROM}\rangle_t/\langle\Nu_\text{FOM}\rangle_t)|\times100)$; time average Reynolds number $\langle \Rey \rangle_t$; and error in $\langle\Rey\rangle_t$ $(\Rey_\text{error}=|(1-\langle\Rey_\text{ROM}\rangle_t/\langle\Rey_\text{FOM}\rangle_t)|\times100)$.}
\label{tab:NuRe_6e6}
\end{table}

\begin{figure*}[hbtp]
\centering
\includegraphics[width=0.75\textwidth]{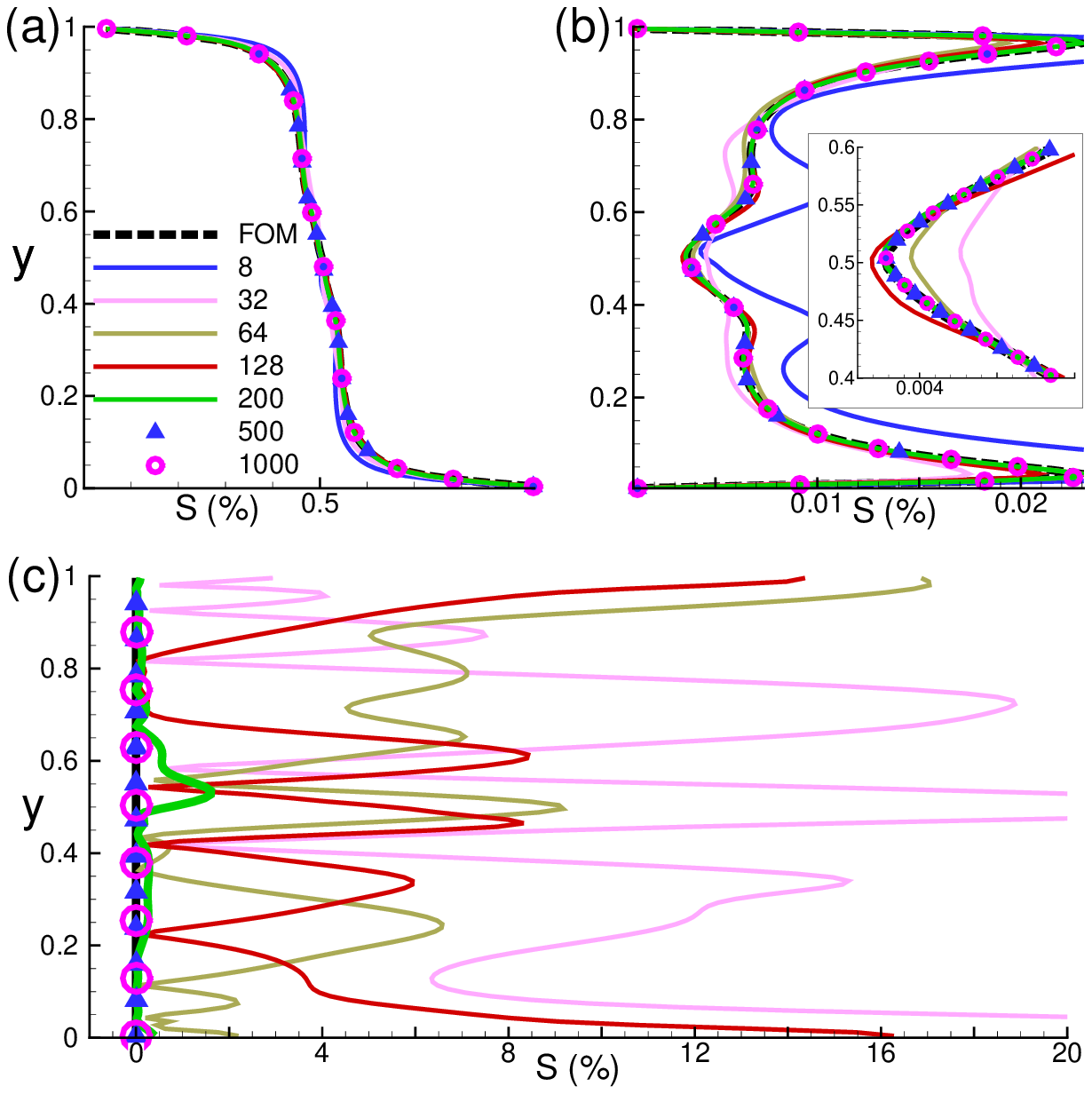}
\caption{For $\Ray=6\times10^6$, vertical profiles of (a) mean temperature $\langle \theta \rangle$, (b) variance of temperature $\sigma_\theta$, and (c) relative error $(S)$ in the ROM profiles of $\sigma_\theta$ with respect to the FOM. Inset shows the zoomed-view.}
\label{fig:meanvar_6e6}
\end{figure*}

\begin{figure*}[ht!]
\centering
     \begin{subfigure}[b]{0.45\textwidth}
         \centering
\includegraphics[width=\textwidth]{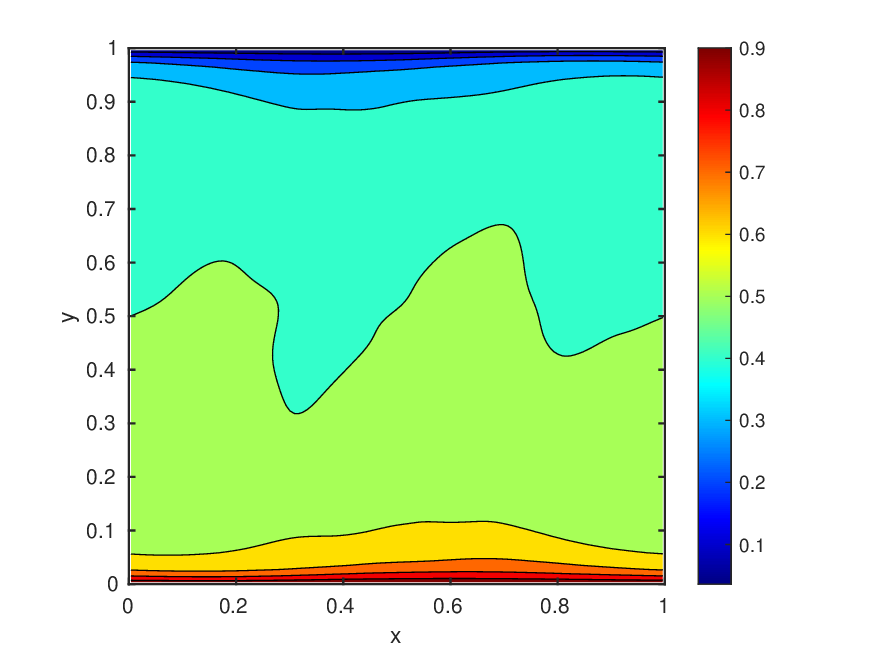}
\caption{FOM}
     \end{subfigure}
     \hfill
     \begin{subfigure}[b]{0.45\textwidth}
         \centering
\includegraphics[width=\textwidth]{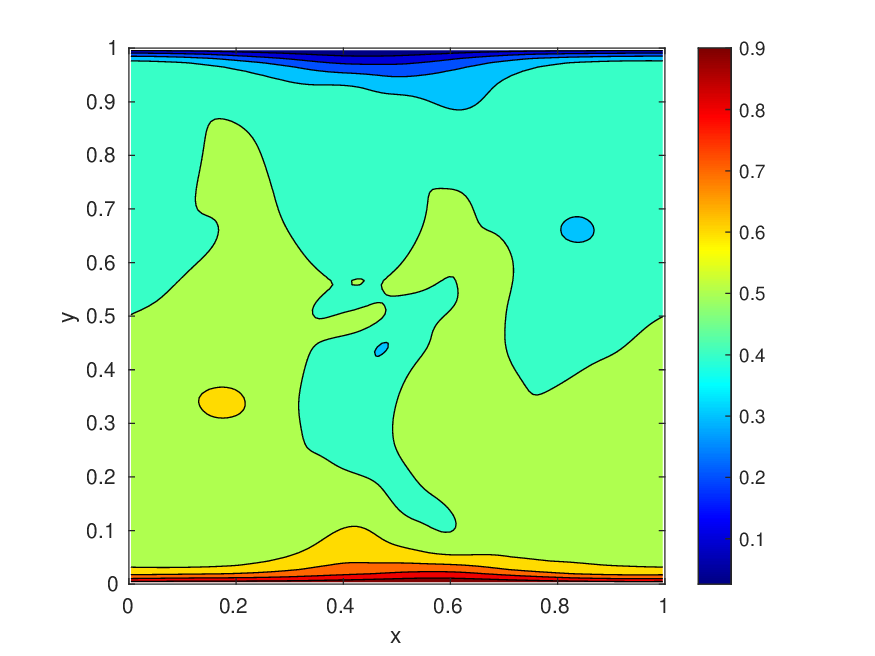}
\caption{ROM with $M= 8$}
     \end{subfigure}
     \hfill
     \begin{subfigure}[b]{0.45\textwidth}
         \centering
\includegraphics[width=\textwidth]{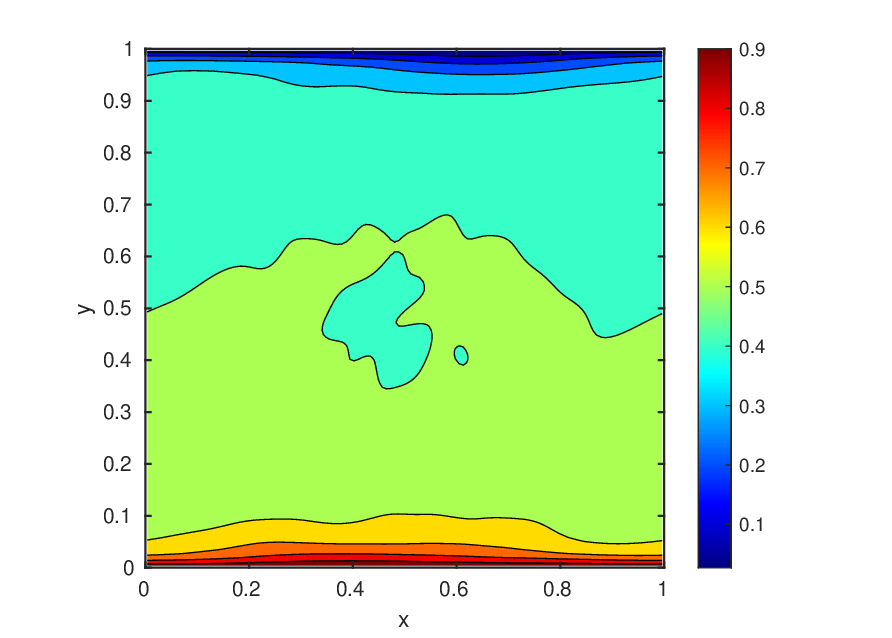}
\caption{ROM with $M= 32$}
     \end{subfigure}
     \hfill
     \begin{subfigure}[b]{0.45\textwidth}
         \centering
\includegraphics[width=\textwidth]{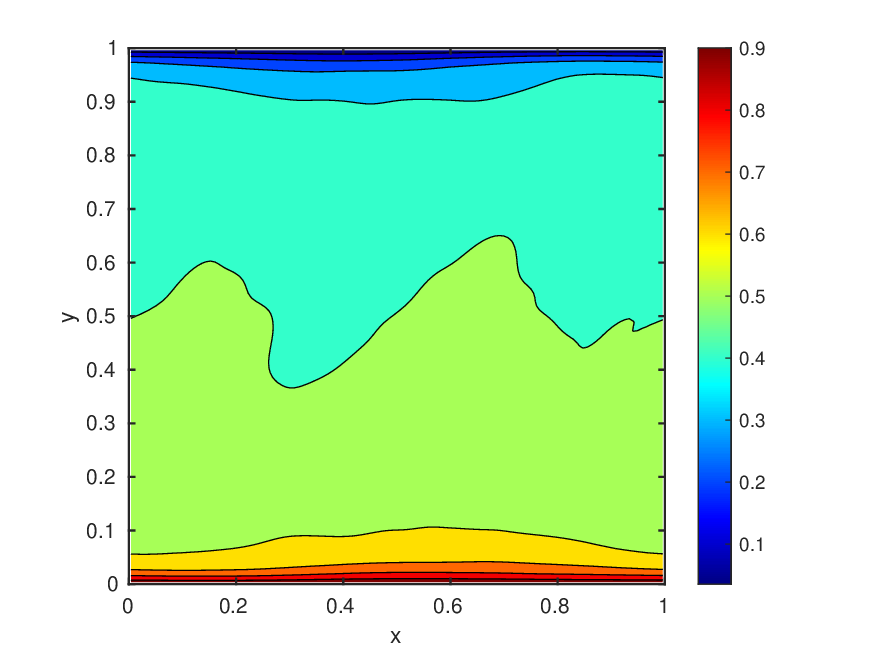}
\caption{ROM with $M= 64$}
     \end{subfigure}
     \hfill
     \begin{subfigure}[b]{0.45\textwidth}
         \centering
\includegraphics[width=\textwidth]{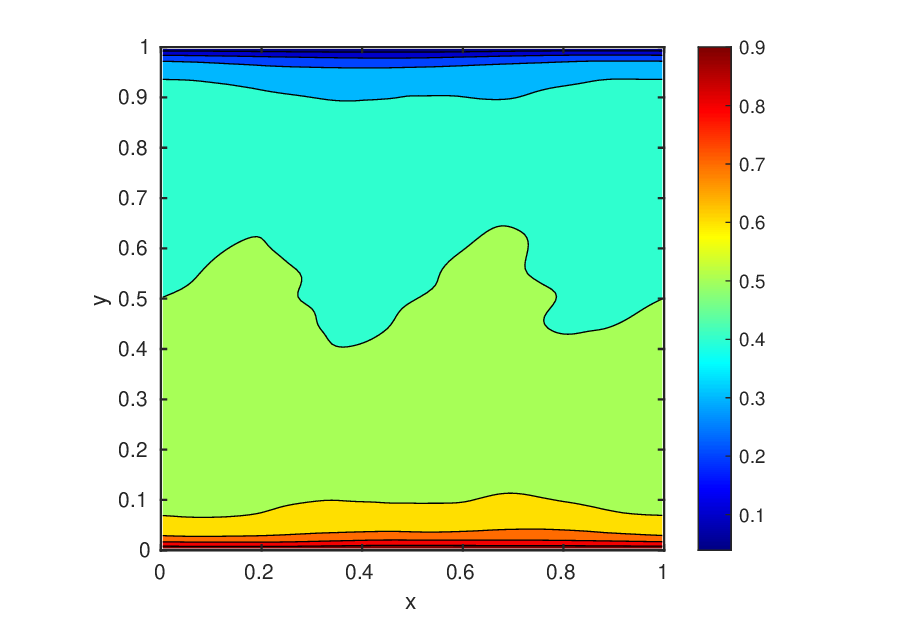}
\caption{ROM with $M= 128$}
     \end{subfigure}
     \hfill
     \begin{subfigure}[b]{0.45\textwidth}
         \centering
\includegraphics[width=\textwidth]{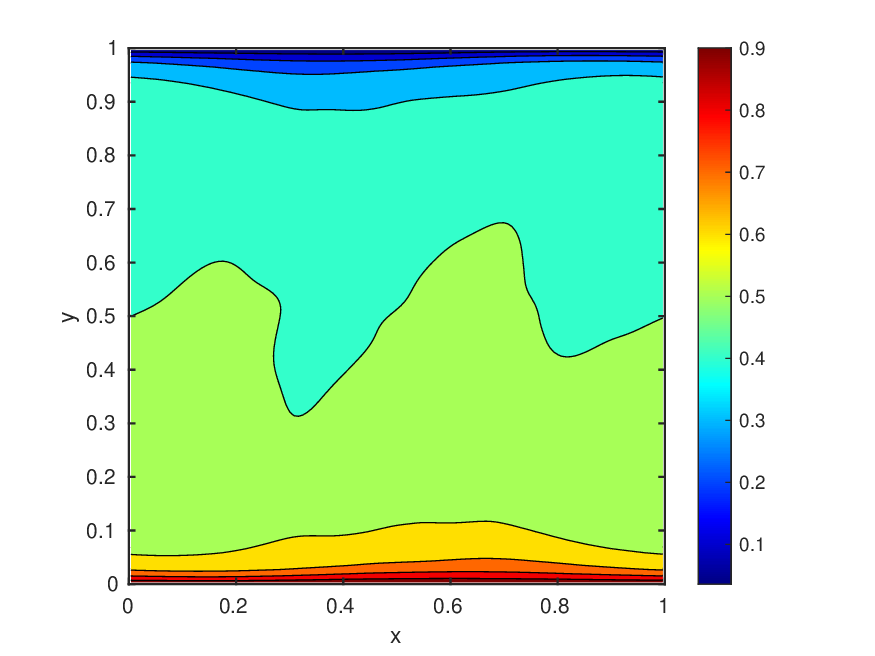}
\caption{ROM with $M= 200$}
     \end{subfigure}
     \hfill
\caption{Comparison of mean temperature field for $\Ray=6\times10^6$. The averaging is carried out for $350$ time units.} 
\label{fig:meantempfield_6e6}
\end{figure*}

We continue to study the temperature statistics in terms of vertical profiles of mean and variance, see Fig.\ \ref{fig:meanvar_6e6}. In line with the observations for the $\Nu$ and $\Rey$ time series, the variance profiles only show  convergence when increasing the number of modes significantly, in this case beyond $M=128$. It is worth mentioning that the relative error converges slowly at certain vertical locations, for instance around $y=0.8$, where the $M=64$ case has a larger error than the $M=32$ case, and much higher $M$ are needed to obtain converged results. On the other hand, the mean temperature profile shows quite clear convergence with increasing number of modes, which is probably due to the fact that it is a `simpler' quantity of interest than the other ones, which involve the temperature gradient or a second-order statistic. In conclusion: it is clearly more difficult to achieve the same level of ROM accuracy in the chaotic case compared to the cases at lower $\Ray$ (steady state and periodic case). This is further confirmed by the mean temperature fields in Fig. \ref{fig:meantempfield_6e6}. It is clearly evident that the mean flow field approaches the FOM as $M$ increases and is highly similar to that of the FOM for $M=200$ (see frame (a) and (f)).

At first sight, not observing convergence up to $M=64$ might seem counter-intuitive, given the fact that the ROM basis already contains more than $99\%$ of the energy, and consequently the best approximation errors will be small. However, the process of dissipation, which is highly important in turbulent flows, happens at scales that carry hardly any energy. Such scales will therefore always be missed by a ROM based on POD (which ranks modes in terms of energy), except when a very large number of modes is included, as shown in our results (for M=200, 500 and 1000). This, however, defeats the very purpose of model reduction. Consequently, for reasonable values of $M$, the delicate balance between energy production, convection and dissipation cannot be accurately represented by the ROM. This is a known issue, see e.g.\ \citep{Wang_2012,Ahmed_2021}, which surfaces here in the study of turbulent natural convection flow. One possible solution is to add a closure model to the ROM equations, like in \citep{Cai_2019}. Our proposed framework, which is stable even without including such a closure model, could form an ideal testbed for developing such closure models.

\section{Conclusions}\label{sec:conclusion}
In the present work, a pressure-free, long-time stable reduced-order model (ROM) is proposed for 2D Rayleigh-B\'{e}nard convection (RBC). The POD-Galerkin ROM is constructed with a `discretize-first, project-later' approach. The discretization on a staggered grid ensures compatibility of the gradient and divergence operators, skew-symmetry of the convective operators (both in momentum transport and temperature transport), and symmetry of the diffusive operators. Upon projection, this results in a pressure-free ROM model that does not suffer from inf-sup stability issues, does not have spurious energy production from the convective terms, and consequently forms an excellent starting point for long-time sampling of statistics in natural convection problems.

In this study, the following control parameters are used: $\Pran=0.71$; $\Ray=10^4, 3\times 10^5,$ and $6\times 10^6$; and a square box with periodic boundary conditions across the lateral walls. The three $\Ray$ cases refer to steady state $(10^4)$, periodic $(3\times 10^5)$ and chaotic $(6\times 10^6)$ cases. Given the sensitivity of the flow dynamics to the initial conditions, two time-independent measures have been used to assess the accuracy of ROMs, i.e., the mean and variance of the vertical temperature profiles. In addition, a comparison of the time-dependent global heat transport properties ($\Nu$ and $\Rey$) is carried out to investigate the accuracy and long-time flow stability. The main observations can be summarize as:
\begin{itemize}
    \item With increasing $\Ray$, the singular-value decay becomes much slower. This can be attributed to the growing importance of the convective motion and the presence of smaller scales. % which plays a significant role in dissipation.
    \item For all cases, the ROMs are stable far beyond their training interval.
    \item For the steady state and periodic flows: increasing the number of modes yields good convergence, where the relative error between FOM and ROM is  below $0.2\%$ in terms of the $\Nu$- and $\Rey$-time series for $M=16$ modes, and below $2\%$ in terms of the mean and variance profiles for $M=32$ modes. 
    \item For the chaotic flow case: the convergence with increasing number of modes is only evident beyond $M=64$, which shows that attaining convergence is relatively difficult in chaotic cases, in contrast to the steady and periodic cases. For example, a ROM with $M=200$ modes is needed to have an error of less than $0.1\%$ in global heat transport properties. The reason for lack of convergence at smaller modes is attributed to the presence of small scale structures, which carry little energy, but are important for the overall energy balance and correct amount of dissipation. 
 \end{itemize}

In summary, the present work has provided a ROM framework to carry out stable long-time sampling in turbulent RBC, which is of great importance to understand heat transport in turbulent natural convection flows. 
%This model could also be used to explore the non-standard variants (inclined and roughness-aided) of RBC, which has recently been studied extensively \cite{Chand_2022,Chand_2021, Chand_2023}. 
The stable nature of the framework forms an excellent starting point to improve its accuracy with closure models that represent the effect of the smallest scales.
\revtwo{Without such closure models, the number of modes required to achieve accurate results in the chaotic regime is too large for the ROM to be computationally efficient. We acknowledge that designing appropriate closure models for the natural convection problem is a difficult task which is beyond the scope of this manuscript. One possible approach is to use the idea of energy conservation (and concomitant stability properties) also in the construction of the closure model. Such closure models are then fully energy-consistent with the remainder of the model, and the long-term stability properties stay intact. We have successfully tested this for one-dimensional problems in \cite{vangastelen2023}, and in future work this can be extended to the Rayleigh-Bénard problem.}
%However, it is also important to note that the current framework (finite volume, staggered grid) is not suitable for non-orthogonal grids, and we would recommend to use instead a discretization suitable for unstructured grids, such as \cite{trias2014}, which satisfies the same properties as our method (skew-symmetry of convection, div-grad compatibility). Also, the proposed ROM is highly intrusive and depends on the FOM details (snapshot matrix, convective and diffusive operators, boundary conditions). In particular, the FOM discretization operators are coded in terms of matrix-vector operations, and building the ROM simply amounts to pre- and post-multiplication these matrices with the ROM basis.}
\section*{Data Availability Statement}
The incompressible Navier-Stokes code is available at \url{https://github.com/bsanderse/INS2D} (Matlab version). A Julia version (without ROM capability) is available from \url{https://github.com/agdestein/IncompressibleNavierStokes.jl}. The data generated in this work is available upon request.\section*{CRediT authorship contribution statement}
\noindent \textbf{K. Chand}: Methodology, Software, Writing - Original Draft, Investigation, Visualization; \textbf{H. Rosenberger}: Methodology, Software, Writing - Review \& Editing, Supervision; \textbf{B. Sanderse}: Conceptualization, Software, Writing - Original Draft, Writing - Review \& Editing, Supervision, Funding acquisition, Project administration\section*{Acknowledgements}
This publication is part of the project "Discretize first, reduce next" (with project number VI.Vidi.193.105 of the research programme NWO Talent Programme Vidi which is (partly) financed by the Dutch Research Council (NWO). We thank the anonymous reviewers for their comments which have helped us to improve the manuscript.

\appendix

\renewcommand\thetable{\arabic{table}}

\section{Grid convergence study}\label{appendix1}

In this section, we provide the details of grid independence study. We have chosen four different grid sizes named as C$_1$, C$_2$, C$_3$, and C$_4$ for coarse to refined grid, respectively. The results of C$_4$ are used as reference to compute errors. To assess the grid independence, we have chosen global $\Nu$ and $\Rey$, averaged over the period in which the flow attains a statistical steady state. Table \ref{tab:gridindependence} shows that the change in the global quantities is less than $1\%$ for both the $\Ray$ cases on grid C$_3$, which is used in section \ref{sec:results}. 

\begin{table}[hbtp]
\centering
\begin{tabular}{*7c}
\toprule
& \multicolumn{3}{c}{$\Ray=3\times10^5$} & \multicolumn{3}{c}{$\Ray=6\times10^6$}\\ \cmidrule(lr){2-4}  \cmidrule(lr){5-7}
Cases & $N_x\times N_y$ & $\langle \Nu \rangle (\Delta \Nu)$ & $\langle\Rey\rangle (\Delta \Rey)$ &  $N_x\times N_y$ & $\langle \Nu \rangle (\Delta \Nu)$ & $\langle\Rey\rangle (\Delta \Rey)$ \\
\midrule 
C$_1$ &  $50\times50$	&  $5.14 (0.58)$	&  $180.95 (0.22)$ &  $80\times80$	    &  $11.63 (0.96)$	&  $1014.93 (0.29)$  \\
C$_2$ &  $64\times64$	&  $5.16 (0.19)$	&  $180.76 (0.11)$ &  $108\times108$	&  $11.70 (1.57)$	&  $1010.75 (0.12)$  \\
C$_3$ &  $80\times80$	&  $5.17 (0.00)$	&  $180.64 (0.04)$ &  $128\times128$	&  $11.48 (0.34) $	&  $1016.44 (0.44)$  \\
C$_4$ &  $96\times96$	&  $5.17 (-)$	&  $180.56 (-)$ &  $144\times144$	&  $11.52 (-) $	&  $1011.98 (-)$  \\     
\bottomrule
\end{tabular}
\caption{Grid convergence study in terms of global $\Nu$ and $\Rey$ for four different grid sizes (C$_1$, C$_2$, and C$_3$) and $\Ray=3\times 10^5$ and $6\times10^6$. $\Delta \Nu=|\Nu_i/\Nu_4-1|)$ where $i=1,2,3,4$, and $(\Delta \Rey=|\Rey_i/\Rey_4-1|)$, where the last case is the reference case. Note that the resolution corresponding to the third case is used in section \ref{sec:results}.}
\label{tab:gridindependence}
\end{table}

\FloatBarrier

\bibliographystyle{elsarticle-num}
\bibliography{rebuttal2/bibgraphy}

\end{document}